\documentclass[authorversion]{acmart}
\AtBeginDocument{%
  \providecommand\BibTeX{{%
    \normalfont B\kern-0.5em{\scshape i\kern-0.25em b}\kern-0.8em\TeX}}}

\usepackage{caption} 
\acmConference[CHI '24]{CHI Workshop on
Human-Notebook Interactions}{May 11--16, 2024}{Honolulu, HI, USA}
\begin{document}

\title[Explainability in JupyterLab and Beyond]{Explainability in JupyterLab and Beyond: Interactive XAI Systems for Integrated and Collaborative Workflows}

\author{Grace Guo}
\orcid{0000-0001-8733-6268}
\affiliation{%
  \institution{Georgia Institute of Technology}
  \streetaddress{North Ave NW}
  \city{Atlanta}
  \state{Georgia}
  \country{USA}
  \postcode{30332}
}
\email{gguo31@gatech.edu}

\author{Dustin Arendt}
\affiliation{%
  \institution{Pacific Northwest National Laboratory}
  \city{Richland}
  \state{Washington}
  \country{USA}
  \postcode{99354}
}
\email{dustin.arendt@pnnl.gov}

\author{Alex Endert}
\orcid{0000-0002-6914-610X}
\affiliation{%
  \institution{Georgia Institute of Technology}
  \streetaddress{North Ave NW}
  \city{Atlanta}
  \state{Georgia}
  \country{USA}
  \postcode{30332}
}
\email{endert@gatech.edu}


\begin{abstract}
    Explainable AI (XAI) tools represent a turn to more human-centered and human-in-the-loop AI approaches that emphasize user needs and perspectives in machine learning model development workflows.
    However, while the majority of ML resources available today are developed for Python computational environments such as JupyterLab and Jupyter Notebook, the same has not been true of interactive XAI systems, which are often still implemented as standalone interfaces.
    In this paper, we address this mismatch by identifying three design patterns for embedding front-end XAI interfaces into Jupyter, namely: 1) One-way communication from Python to JavaScript, 2) Two-way data synchronization, and 3) Bi-directional callbacks.
    We also provide an open-source toolkit, bonXAI, that demonstrates how each design pattern might be used to build interactive XAI tools for a Pytorch text classification workflow.
    Finally, we conclude with a discussion of best practices and open questions.
    Our aims for this paper are to discuss how interactive XAI tools might be developed for computational notebooks, and how they can better integrate into existing model development workflows to support more collaborative, human-centered AI.
\end{abstract}



\begin{teaserfigure}
  \includegraphics[width=\textwidth]{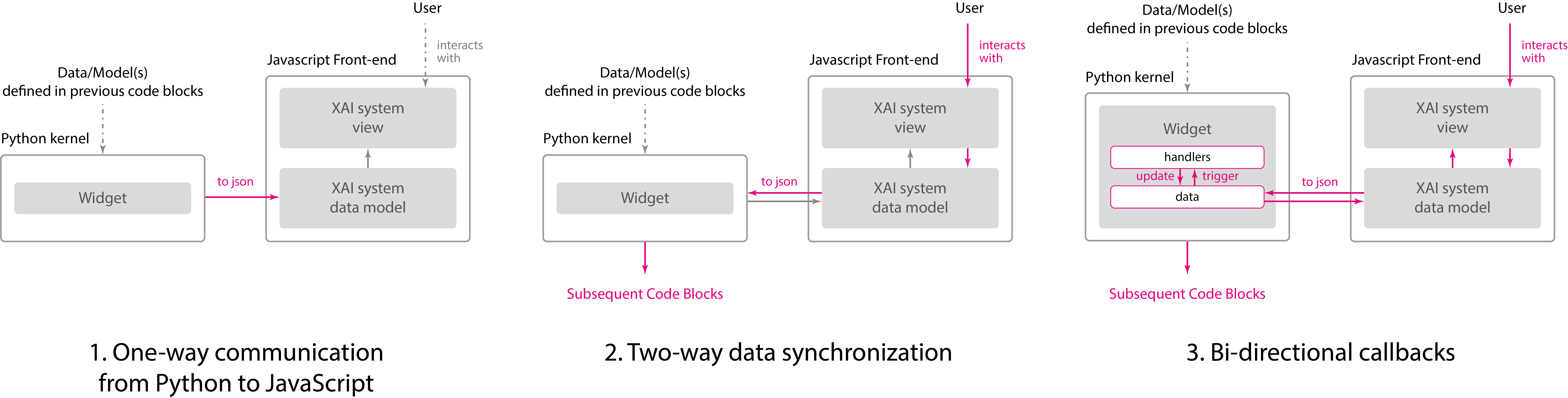}
  \caption{Three design patterns for embedding XAI systems into JupyterLab notebooks.}
  \Description{}
  \label{fig:teaser}
\end{teaserfigure}

\received{20 February 2007}
\received[revised]{12 March 2009}
\received[accepted]{5 June 2009}

\maketitle

\section{Introduction}
The recent advancements in machine learning (ML) and artificial intelligence (AI) have led to applications of AI models in diverse contexts ranging from medicine and healthcare \cite{sermesant_applications_2021, van_assen_artificial_2023, rouzrokh_machine_2023} to education and fraud detection \cite{a2020neva, leite2017eva, corbett1997intelligent}.
But while these AI models can mimic or exceed human proficiencies, they often lack transparency and can be inscrutable to ML developers and end-users.
This lack of transparency may lead to unintended harms that persist even when the models themselves are removed \cite{weidinger2021ethical, ehsan2022algorithmic, hagerty2019global, ackerman2000intellectual, ehsan2021expanding, ehsan2023charting}.
To mitigate these consequences, various explainable AI (XAI) tools and systems have thus been developed that aim to open the ``black box'' of AI such that developers, end-users, and stakeholders can better inspect, interpret, and refine model performance \cite{arrieta2020explainable, ehsan2022human, ehsan2020human, ehsan2021explainable, ehsan2021operationalizing}.
By enhancing the interpretability and transparency of models, XAI represents a turn to more human-centered and human-in-the-loop AI approaches that better address user needs and perspectives.

Within XAI, interactivity has emerged as an important factor for effective explanations \cite{miller2019explanation, langer2021we, arya2019one}.
Interactivity helps support human-centered and human-in-the-loop AI approaches by facilitating conversation between user and model \cite{adadi2018peeking}, and the use of such interactive XAI systems has been found to improve prediction outcomes by allowing end-users to refine models based on their domain expertise or contextual knowledge \cite{bertrand2023selective}.
To date, these XAI tools are typically implemented as in-browser applications, largely due to the wide variety of JavaScript toolkits and libraries that support richly interactive browser-based interfaces and visualizations.

In laying out his vision for literate programming, Donald Knuth proposed that \textit{``Instead of imagining that our main task is to instruct a computer what to do, let us concentrate rather on explaining to human beings what we want a computer to do''} \cite{knuth1984literate}.
This aligns closely with the goals of many human-centered AI and XAI approaches, yet there is a mismatch in how XAI tools and ML models are built and used.
While JavaScript browser applications are common for XAI, the core of resources available for ML are developed for Python -- from Pytorch \cite{PaszkePyTorchAnImperative2019} and Tensorflow \cite{tensorflow2015whitepaper} to platforms such as Hugging Face\footnote{https://huggingface.co/} that provide a central repository of ML resources and pipelines.
The dominance of these tools means that Python programming environments, such as JupyterLab and Jupyter Notebook\footnote{https://jupyter.org/}, have also become key components of ML workflows.
However, this mismatch in usage environments between XAI tools and ML models means that model explanations often cannot be presented \textit{in context} of how the model is built and used.
Developers who wish to adopt XAI in their work typically need two distributed processes running on the same machine, and must individually manage the challenges of data communication and state synchronization between model and explanation.

Solutions to these challenges have been influenced by available tools and their ease of use.
One typical approach combines a Python server back-end with a browser front-end written in JavaScript \cite{ahn2022tribe, kouvela2020bot, sovrano2021philosophy}.
However, this solution does not support the integration of XAI tools into notebook environments.
Other solutions -- such as IPyWidgets \cite{JupyterWidgets}, NOVA \cite{wang2022nova} and anywidget \cite{Manz_anywidget} -- have been proposed that aim to embed XAI interfaces directly into notebook cells.
However, some of these frameworks are limited to one-directional data communication.
NOVA, for example, does not propagate user inputs back to the Python kernel, which creates challenges for incorporating user feedback into model development workflows.

In our prior work \cite{guo2021vaine, guo2023causalvis}, we have addressed these challenges by adapting the existing IPyWidgets framework to build interactive XAI systems with two-way communication such that data/states are synchronized between the front-end interface and back-end Python kernel.
This implementation meant that user inputs could be accessed in subsequent notebook cells and used to trigger callback functions to run data analytics and model pipelines.
For data analysts and model developers, these embedded XAI systems facilitated better communication of the model development process by presenting explanations in context of the code.
The explanations also helped them engage end-users in tasks of data selection, outcome evaluation, and iterative model refinement that would otherwise require programming expertise.


In this paper, we reflect on our prior work to identify three design patterns for how front-end XAI interfaces can be embedded into Jupyter, namely: 1) One-way communication from Python to JavaScript, 2) Two-way data synchronization, and 3) Bi-directional callbacks (Fig. \ref{fig:teaser}).
We also provide an open-source toolkit, bonXAI, that provides examples of how each design pattern might be used to build interactive XAI systems for a Pytorch text classification workflow.
Finally, we outline some best practices, design guidelines, and open questions for future research.
In sum, our aims for this paper are to discuss how interactive systems can be developed for computational notebooks, and how XAI tools can better integrate into existing model development workflows to support collaborative, human-centered AI.

\section{Related Work}

In this section, we provide an overview of interactive XAI approaches and systems developed in prior work.
This overview highlights how interactivity is both crucial to the effectiveness of AI explanations, while, at the same time, many explanations are still presented independently of the model being explained.
Due to the expansive landscape of current research, we cannot enumerate all XAI systems here.
For recent surveys refer to \cite{adadi2018peeking} and \cite{bertrand2023selective}.

This section will also provide a summary of libraries for building interactive widgets in computational notebooks, particularly for the Jupyter notebook environments.
To date, there are two notebook versions offered by Project Jupyter: the more recent JupyterLab interface, and the classic Jupyter Notebook\footnote{Project Jupyter recommends using JupyterLab since the classic Notebook is being deprecated. However, both environments remain in use, and our work supports both.}.
In the rest of this paper and unless otherwise stated, we use Jupyter to refer to both environments.
As our own prior works \cite{guo2021vaine, guo2023causalvis} have primarily used Python and Jupyter, they are also the focus of this overview.
However, other programming languages and computational environments exist, such as Google Colab\footnote{https://colab.research.google.com/} (an in-browser version of Jupyter), that have similarly seen widespread adoption by developers. We discuss them in relation to embedded XAI systems in Section \ref{beyond_jupyterlab}.

\subsection{Interactive XAI Systems} \label{interactive_xai}

AI models have often been described as ``black boxes'' due to their lack of transparency and interpretability.
Explainability tools, also referred to as XAI systems, have thus been developed to help open these ``black boxes'' and explain how AI models work.
In recent years, interactivity has emerged as one crucial factor influencing the effectiveness of these explanations.
In a notable 2018 survey, Adadi and Berrada propose that ``explainability can only happen [through] interaction between human and machine'' \cite{adadi2018peeking}.
This has since been supported by work from other researchers arguing from psychological, philosophical and social perspectives that interactivity is an important factor for good explanations \cite{miller2019explanation, langer2021we, arya2019one}.
And in a 2023 paper looking at 48 evaluations of interactive XAI systems, Bertrand et al. found empirical evidence that ``interactive explanations improve perceived usefulness and performance of the human+AI team'' \cite{bertrand2023selective}.

Beyond improving outcomes, prior studies have also found that the use of XAI systems can better support human-centered AI approaches.
For ML model developers, a wide range of tools exist to support tasks from data pre-processing to model evaluation and refinement \cite{arrieta2020explainable, arya2019one, sperrle2021survey}.
In contrast, far fewer XAI systems have been developed for or evaluated with end-users who are affected by the models \cite{bhatt2020explainable, hong2020human, liao2020questioning, sperrle2021survey}.
For end-users, explanations tend to focus on providing insight into model outcomes in specific usage domains, such as healthcare \cite{cai2019hello, cai2019human} and education \cite{fiok2022explainable, khosravi2022explainable}.
However, in a recent study of end-user needs and expectations by Kim et al., the authors found that end-users, like model developers, wanted XAI that do more than explain model outcomes.
Instead, participants in the study expressed a desire for tools that engage them in model development, helping them better calibrate trust, improve task skills, supply better inputs, and give constructive feedback to developers \cite{kim2023help}.
The authors went on to propose an set of recommendations to guide the implementation of interactive XAI tools for end-users to actively participate in model development workflows, but how these recommendations should be realized remains open to study.

Taken together, these prior works lay out a compelling case for the effectiveness of interactive XAI tools, as well as their potential as a means to engage end-users without ML expertise in the development and application of ML models.
In this paper, we explore how interactive XAI systems can be \textit{embedded into computational environments} as a means for AI model development workflows to support collaboration between users of different backgrounds \cite{kim2023help, ehsan2021explainable, liao2021human}, and seamlessly integrate user feedback and judgements into model development processes.

\subsection{Computational Notebooks and the IPyWidgets Framework} \label{ipywidgets}

Donald Knuth first laid out the vision of ``literate programming'' in his 1984 paper, which posits that programs are works of literature for \textit{humans} rather than machines and should be documented as such \cite{knuth1984literate}.
This vision has since been developed and extended into a range of computational notebook environments that allow developers to display executable code with code outputs, multimedia text, images, video, audio, and interactive interfaces in a single document.
These notebook environments help communicate the work of a program for human audiences, supporting collaboration, reproducibility, sharing, and communication, particularly in data science workflows \cite{kluyver2016jupyter, schwab2000making, shen2014interactive}.
The benefits of working in notebook environments has since led to widespread adoption \cite{kaggle}, and has motivated research into how they might be extended into contexts of non-linear analysis \cite{wang2022stickyland, wu2020b2, weinman2021fork}, workflow recommendation \cite{raghunandan2024lodestar}, automatic narrative generation \cite{kery2018story, wang2023slide4n, zheng2022telling, li2023notable, kang2021toonnote}, and fluid transitions between code and graphical interfaces \cite{kery2020mage}.

In line with many of these prior studies, our work looks at the Jupyter computational environments, and specifically the IPyWidgets framework \cite{JupyterWidgets}.
The framework was designed to help embed interactive front-end interfaces into notebook cells by providing a communications layer that sends data in JSON format between the front-end interface and the back-end Python kernel.
This communications layer ensures that the data (or states) are synchronized within a Jupyter \textit{widget}.
On widget instantiation, the front-end and the back-end store copies of the same data/state.
When users interact with the interface, any changes to the data can be automatically propagated back to the kernel using \textit{traitlets}, allowing the new data to be accessible in the notebook.
Handlers can also be registered to these data attributes such that any change triggers a callback function that can be used to automatically run some data analysis or, in the ML context, a pre-specified model.
In this paper, we leverage this IPyWidgets framework to define three design patterns for how interactive XAI systems can be embedded into Jupyter workflows.

\subsection{Interactive XAI Widgets for Jupyter}

Python libraries and computational environments make up the vast majority of current AI and ML tools \cite{kaggle}.
From libraries such as scikit-learn \cite{scikitlearn}, Pytorch \cite{PaszkePyTorchAnImperative2019}, Keras \cite{chollet2015keras} and Tensorflow \cite{tensorflow2015whitepaper}, to platforms such as Hugging Face, these resources have built up a robust Python ecosystem for model development and use.
Consequently, Python computational environments, such as Jupyter and Google Colab, have also become some of the most common environments for machine learning.
All libraries listed above, for example, include some mention of Jupyter, Colab, or ``notebooks'' in their examples and tutorials.

However, while Python and Juptyer have become the dominant tools for machine learning development, this has not been the case for XAI systems, particularly interactive XAI systems.
Consider the tools surveyed by Bertrand et al. \cite{bertrand2023selective}.
One typical architecture implements the system interface using a JavaScript front-end while a Python server back-end provides the ML model being explained \cite{ahn2022tribe, kouvela2020bot, sovrano2021philosophy}.
This approach is useful, but requires additional work to set up and connect the server and the interface.
Other implementations have sought to run the XAI system interface and the ML models from a single environment.
For example, Ross et al. were able to implement models directly in-browser by first converting them to Tensorflow.js \cite{tensorflow2015whitepaper, ross2021evaluating}, while Spinner et al. build their XAI visualizations as Tensorboard\footnote{https://www.tensorflow.org/tensorboard} plugins that can be used by any developer designing and training a TensorFlow model \cite{spinner2019explainer}.
In the latter example, the adoption of Tensorboard helps provide a more seamless integration of XAI into existing model development processes for users who are already working with the Tensorflow library.
However, since Tensorboard visualizations are hosted in a separate browser window, this set-up still requires frequent context switching between the coding environment and the in-browser explanations.
This can also make it challenging to obtain an end-to-end overview of model development with the XAI systems inserted in an ordered, linear narrative.
More critically, when we look at the 48 papers included in Bertrand et al.'s survey \cite{bertrand2023selective}, most do not provide any detail about how the interactive XAI systems are implemented and used.
System screenshots also typically exclude details about computational environments or the expected usage context of the tool, which can result in barriers to adoption when ML developers are not shown how they can quickly incorporate these XAI systems into their work.

More commonly, toolkits built for Jupyter can be used to explain ML models.
These include general purpose visualization libraries -- such as matplotlib \cite{Hunter2007}, seaborn \cite{Waskom2021}, and Altair \cite{VanderPlas2018} -- and specialized XAI tools -- such as Captum \cite{kokhlikyan2020captum}, Class Activation Mapping methods \cite{jacobgilpytorchcam}, ELI5 \cite{ELI5}, InterpretML \cite{nori2019interpretml}, LIME \cite{lime} and SHAP \cite{shap}.
Of these, the majority are static visualizations.
While some libraries include interactivity (e.g. Altair), they tend to be visualization-centric and are not designed to support specific XAI tasks.
To address these limitations, frameworks, such as NOVA \cite{wang2022nova}, have been built to help embed interactive JavaScript interfaces into notebooks.
These frameworks simplify the process of wrapping complex XAI systems into notebook widgets, allowing ML developers to rapidly add explanations to their existing workflows and obtain insights about the data or model.
However, these widgets typically only support the one-way communication from the notebook to the JavaScript interface, and changes to the data/state caused by user interactions with the interface can neither be propagated back to the Python kernel nor accessed in subsequent notebook cells.
This is a limitation for interactive XAI systems (such as the ones discussed in Section \ref{interactive_xai}) where user feedback may be necessary to evaluate, refine or update a model.
In particular, two-way data communication and state synchronization can be essential in collaborative situations where end-users without programming expertise may benefit from inspecting models and providing input through an interactive front-end interface.

In this paper, we build on existing widget frameworks \cite{JupyterWidgets, wang2022nova, Manz_anywidget} and our own prior work \cite{guo2021vaine, guo2023causalvis} to define three design patterns of how interactive XAI systems can be embedded into ML development workflows in Jupyter.
These design patterns provide a basic template for displaying the interfaces, synchronizing data, automatically propagating user input back to the kernel, and triggering callback functions.
Taken together, this paper aims to provide an exploration of how interactive XAI systems can be more tightly integrated into model development environments to better support human-centered and human-in-the-loop AI approaches.

\section{Three Design Patterns for XAI in Jupyter} \label{designpatterns}

We use the IPyWidgets framework (see Section \ref{ipywidgets}) to embed interactive XAI systems into Jupyter.
A widget can be instantiated in a notebook code block with relevant arguments (parameters) in the manner of \verb|widget(*args)|.
On instantiation, the JavaScript front-end of the widget will be rendered in an output notebook cell.
From existing XAI systems and our own prior work, we identified three design patterns for the IPyWidgets framework: 1) One-way communication from Python to JavaScript, 2) Two-way data synchronization, and 3) Bi-directional callbacks (Fig. \ref{fig:teaser}).

To demonstrate how each design pattern might be implemented, we also provide a gallery of example XAI widgets in our supplemental toolkit, bonXAI\footnote{https://github.com/gracegsy/bonXAI}.
This toolkit includes three widgets: 1) DataExplorer, 2) DataSelector, and 3) InferenceExplorer that correspond to the three design patterns.
They are developed for the Hugging Face PyTorch text classification workflow \cite{HuggingFacetextclassification, PaszkePyTorchAnImperative2019}, but can be generalized to other textual data exploration, selection and inference tasks.
BonXAI uses React\footnote{https://reactjs.org/} and the IPyWidgets framework, but readers may consider using other implementations when building XAI tools for Jupyter.

\begin{figure*}[ht]
  \includegraphics[width=\textwidth]{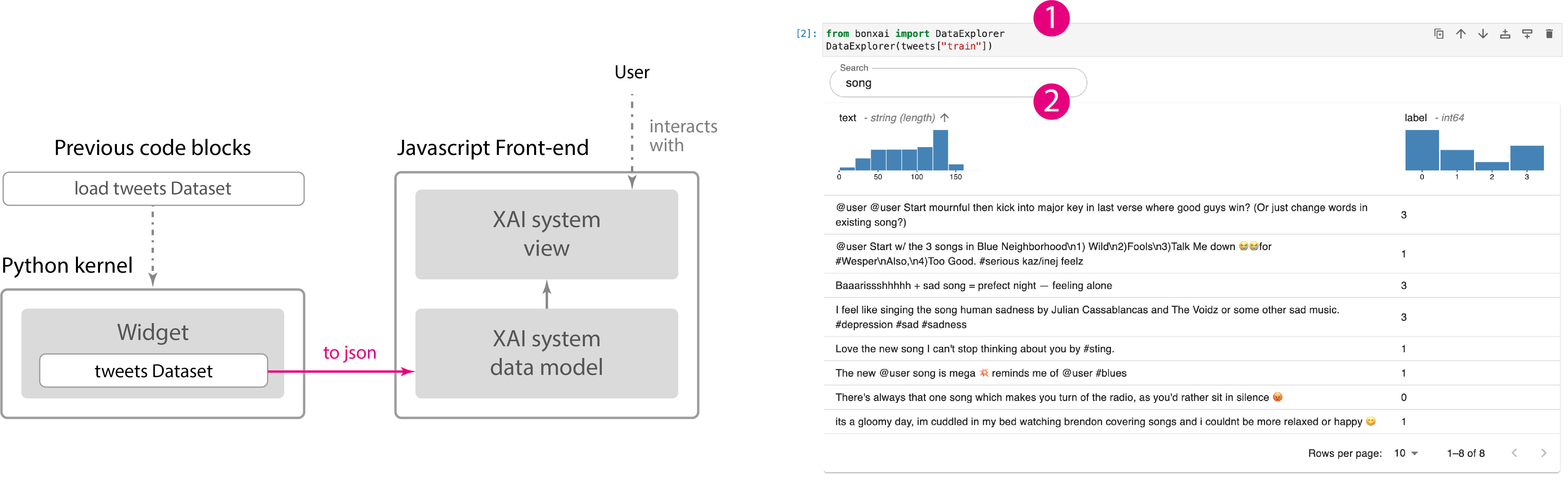}
  \caption{The DataExplorer widget is an example of design pattern 1. This type of XAI system provides a display based on a given a model and/or dataset as input. \textcircled{1} The DataExplorer widget is instantiated with some data that is converted to JSON and sent to the front-end. \textcircled{2} While users can interact with the display, their interactions do not change data attributes on the back-end Python kernel.}
  \Description{}
  \label{fig:designpattern_1}
\end{figure*}

\subsection{Design Pattern 1: One-way communication from
Python to JavaScript}

The first design pattern for embedded XAI systems involves basic JavaScript displays in notebook cells.
Such systems are typically instantiated by passing in some data that is then displayed in the front-end view of the widget.
This design pattern is widely used in many XAI toolkits for Jupyter -- such as Captum \cite{kokhlikyan2020captum}, Class Activation Mapping methods \cite{jacobgilpytorchcam}, ELI5 \cite{ELI5}, InterpretML \cite{nori2019interpretml}, LIME \cite{lime} and SHAP \cite{shap}.
However, the majority of these explanations are static displays.
Even in cases where the user may be able to interact and explore the data/model further, the results of their interactions are not propagated back to the Python kernel \cite{guo2021vaine, alammar2021ecco}.
A key characteristic of this design pattern is this one-way communication of information from notebook to XAI system.
User interactions (if any) are contained in the front-end interface, and cannot be accessed in subsequent code blocks.

In our demonstration library, we provide an example of this design pattern with the DataExplorer widget (Fig. \ref{fig:designpattern_1}).
We can pass a Hugging Face text Dataset to this widget, which is displayed in an interactive table for exploration.
Users can search for substrings, or filter by string length and other variables.
However, these interactions do not change any attributes on the Python back-end, and the filtered subset is not accessible in subsequent notebook code blocks.

To date, this design pattern is, by and large, the most common approach for embedding XAI systems into Jupyter.
A variety of frameworks, such as NOVA \cite{wang2022nova}, have also been developed to simplify implementation by automating the process of wrapping front-end JavaScript into widgets.
The resulting toolkits are well suited for post-hoc exploring, understanding, and communicating details about a dataset, model, or training process.
However, since interactions are not propagated back to the kernel, these XAI systems may be insufficient for human-in-the-loop tasks that rely on user input.


\begin{figure*}[ht]
  \includegraphics[width=\textwidth]{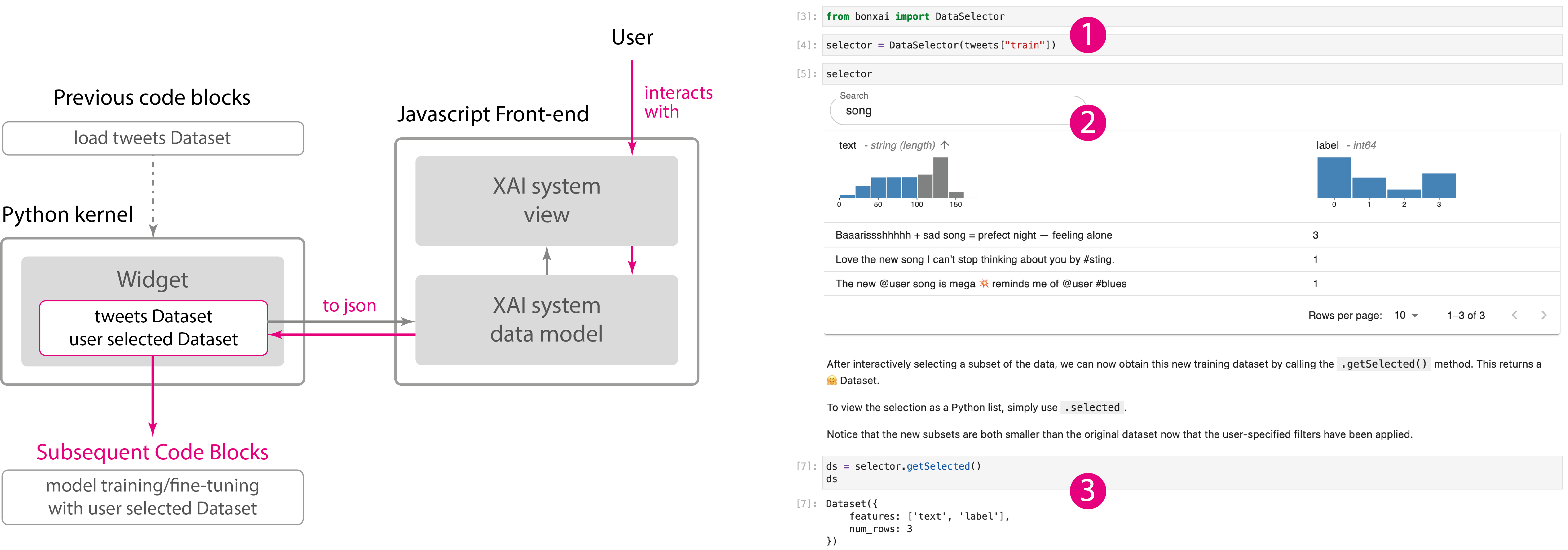}
  \caption{The DataSelector widget is an example of design pattern 2, which synchronizes the data/states between the front-end and back-end, allowing user inputs to be propagated back to the notebook. \textcircled{1} As before, the DataSelector widget is instantiated with some data that is converted to JSON and sent to the front-end. \textcircled{2} Users can interactively explore the data set by searching or applying filters. \textcircled{3} The filtered data set from user interactions can be accessed in subsequent notebook cells/code blocks.}
  \Description{}
  \label{fig:designpattern_2}
\end{figure*}

\subsection{Design Pattern 2: Two-way data synchronization} \label{dataselector}

The second design pattern supports the two-way communication of data.
In such XAI systems, when users interact with front-end input components, the results of their interactions are propagated back to the kernel to update the relevant data structures and/or variables.
In this way, data/states are synchronized between the interface and the kernel.
User input can thus be accessed in subsequent notebook code blocks and applied back to the data analysis process.
To the best of our knowledge, this two-way communication must be manually implemented using a framework, such as IPyWidgets or anywidget, and no libraries or tools have been developed to automate the process as yet.

Increasingly, researchers have called for the engagement of domain experts and end-users in order to develop machine learning methods that better incorporate domain knowledge and user judgements \cite{doshi2017towards}.
However, extracting this knowledge can be challenging since domain experts and end-users may not be familiar enough with ML or programming to input their feedback directly in code.
In such situations, a front-end system embedded into Jupyter can potentially better facilitate collaborative efforts between developers and experts.
The front-end interface is an intuitive way for domain experts and end-users to interactively provide their domain knowledge, while the propagation of this input back to the notebook means that developers can quickly incorporate updates into subsequent code blocks and ML pipelines.

In one of our prior studies \cite{guo2023causalvis}, we applied this design pattern in three embedded interactive visualizations for causal inference analysis.
Using these visualizations, causal inference analysts and domain experts can collaboratively edit the outputs of causal discovery algorithms, discuss the appropriate variables to use in analysis, inspect the data instances to include or exclude, and explore subgroup differences in estimated treatment outcomes.
We provide another demonstration of this design pattern with the DataSelector module in our example library (Fig. \ref{fig:designpattern_2}).
This module is similar to the DataExplorer, however, in addition to allowing users to interactively explore their data, the filtered subset is also sent back to the Python kernel and can be accessed in subsequent notebook code blocks as a Hugging Face Dataset.
Like any other Dataset provided by the Hugging Face Datasets library, this filtered subset can also be used for model training and fine-tuning.

More generally, we see potential for embedded XAI systems of this design pattern to be adapted for ``data explanations'' \cite{arya2019one} and other data-centric tasks.
These are tasks that require collaboration or expert judgements, such as data labeling and cleaning \cite{roh2019survey, whang2023data, whang2020data, krishnan2016activeclean, krishnan2016towards} or curating human feedback for model fine-tuning using RLHF \cite{lambert2022illustrating, knox2008tamer, macglashan2017interactive, christiano2017deep, warnell2018deep, wirth2017survey}.

\begin{figure*}[h]
  \includegraphics[width=\textwidth]{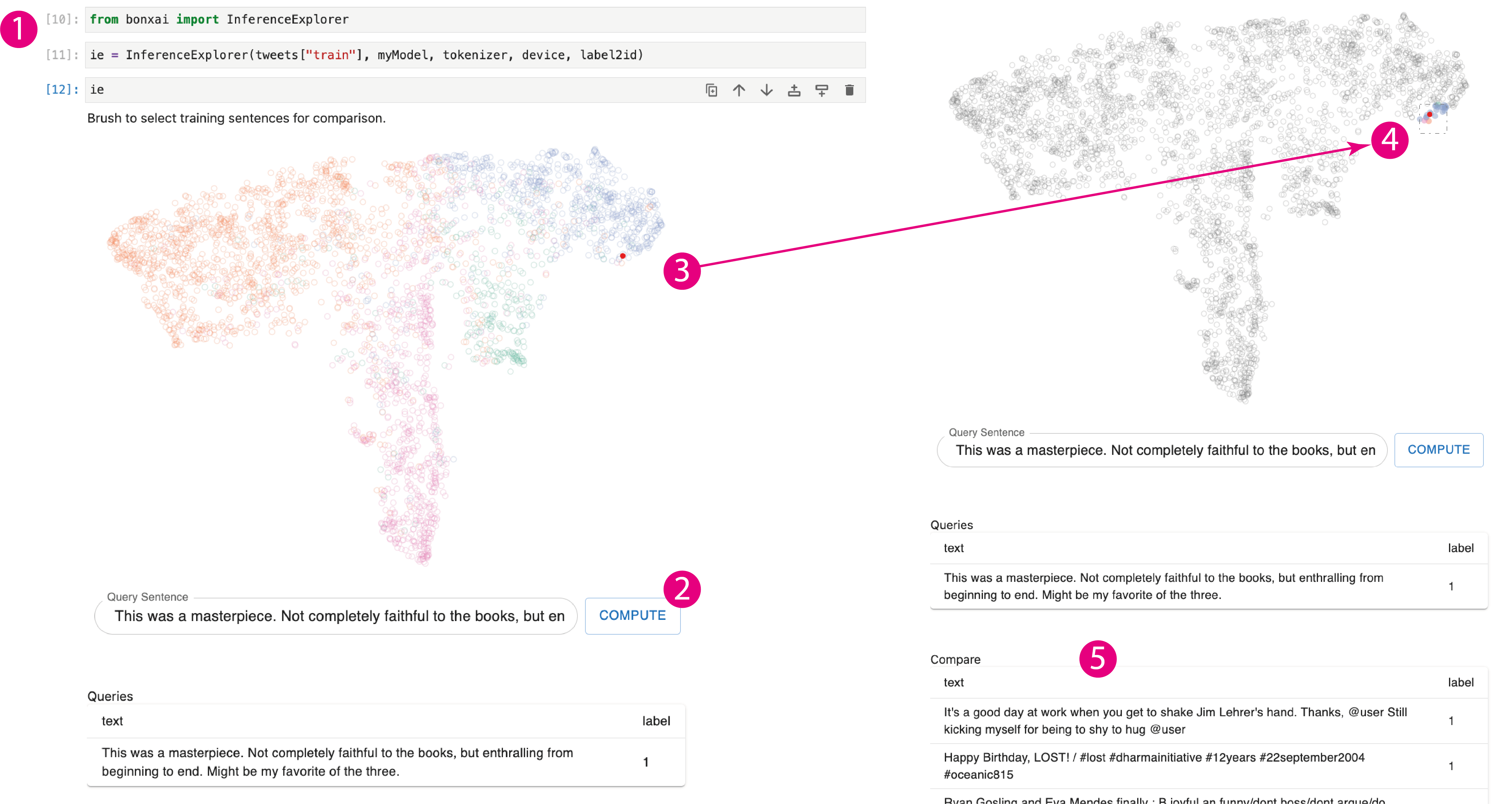}
  \caption{The InferenceExplorer module demonstrates how Design Pattern 3 might be applied. \textcircled{1} Like in previous examples, data is converted to JSON and sent to the front-end. In this widget, we also provide a pre-trained model. Since a model cannot be converted to JSON format, the model is simply registered as a widget attribute in the Python back-end. The widget computes and visualizes the UMAP embedding of all text in the data set. \textcircled{2} Users can interactively input some new text for inference. Its label, predicted by the pre-trained model on the Python back-end, is displayed in the table below. \textcircled{3} The position of this new text is also computed and added to the UMAP embedding visualization as a red dot. \textcircled{4} To \textit{explain} why their input was given a particular label, users can brush and select its nearest neighbors in the visualization. \textcircled{5} These selected neighbors are listed in a separate table. We can see that tonally, they resemble the user input text, and have the same label.}
  \Description{}
  \label{fig:inference_explorer}
\end{figure*}

\begin{figure}
  \includegraphics[width=0.5\columnwidth]{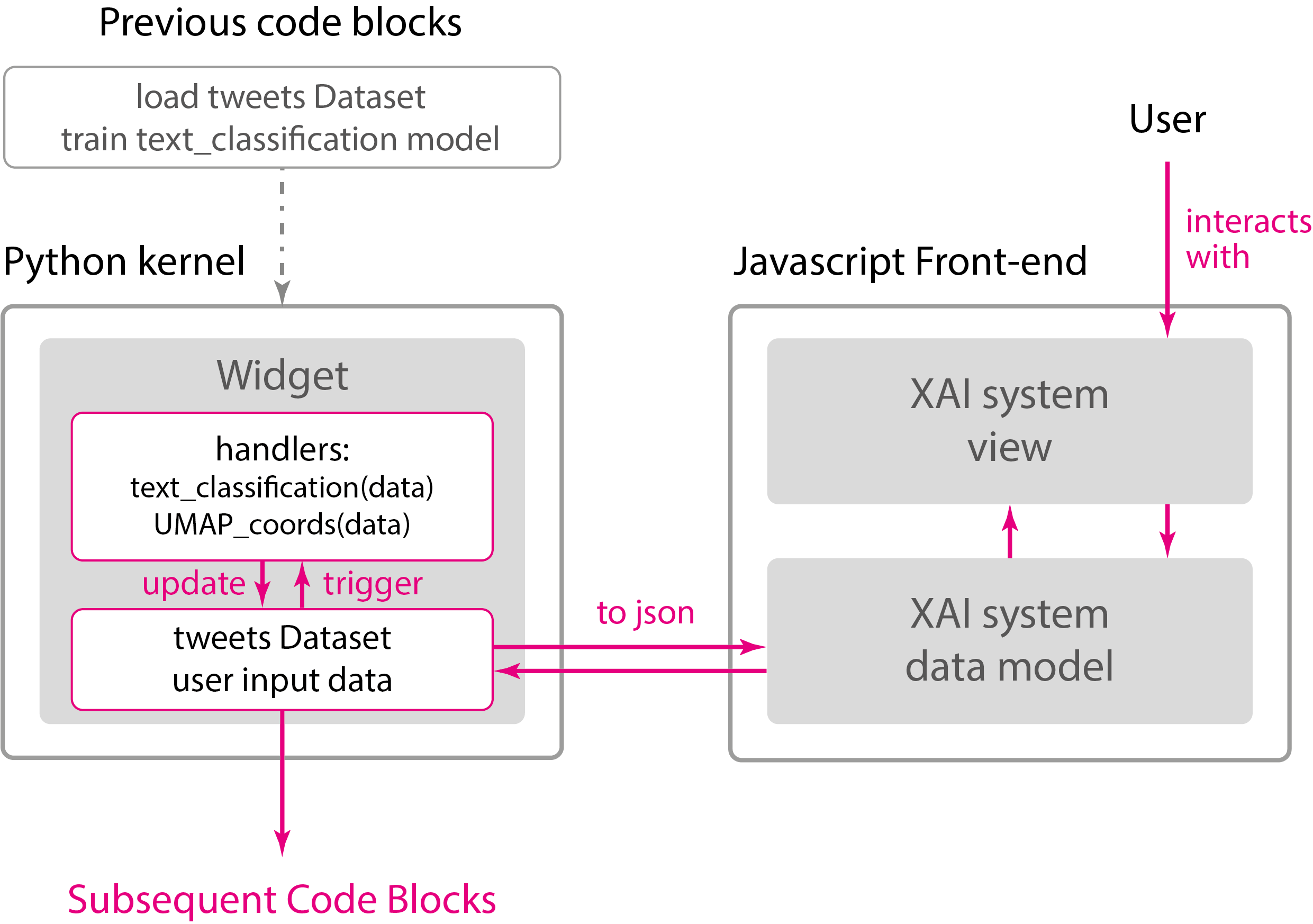}
  \captionsetup{width=.5\columnwidth}
  \caption{The InferenceExplorer module applies the third design pattern to provide bi-directional callbacks. When users interact with input components, their interactions update the data/state on the Python kernel back-end. These changes automatically trigger callback functions for data processing, model inferencing, and updating the front-end display.}
  \Description{}
  \label{fig:designpattern_3}
\end{figure}

\subsection{Design Pattern 3: Bi-directional callbacks}

The third design pattern of embedded XAI systems extends the two-way communication such that changes to the data/state also trigger callback functions in the kernel that execute additional data processing tasks.
The results of these callback functions are passed back to the front-end interface and used to update the display, thus ensuring that any new data or user input is dynamically reflected in the XAI tool.
This design pattern most closely realizes the goals of human-in-the-loop AI approaches by allowing end-users to engage in tasks that directly affect model development and use, such as prompt tuning, inference validation, model refinement and others.

In our demonstration library, we have provided an example of how one such XAI system can be embedded into Jupyter (Fig. \ref{fig:inference_explorer} and \ref{fig:designpattern_3}).
Similar to the previous examples, the InferenceExplorer widget is instantiated with a training data set.
For each line of text in the data set, the widget first gets its embedding, then performs a UMAP dimensionality reduction \cite{mcinnes2018umapsoftware} to obtain 2D-coordinates for all instances.
These coordinates are visualized to provide an overview of the data set, and can be used to determine how well the model distinguishes the different labels.

This InferenceExplorer widget also applies the third design pattern to support bi-directional callbacks.
One of the parameters required to instantiate this widget is a pre-trained model.
When users interact with the front-end interface, they can choose to input new sentences for inference.
This input is propagated back to the widget, updating an internal data attribute.
In addition to this two-way data communication and synchronization, a \textit{handler} automatically triggers a callback function that uses the pre-trained model to get the predicted label for the new sentence input.
This handler function also computes 2D coordinates for the sentence.
Finally, the text, predicted label, and umap coordinates are sent back to the front-end to be added to the visualization as a red dot.

By adopting bi-directional callbacks in our InferenceExplorer widget example, the XAI system closes the loop between user input, model inference, and visual explanation.
Users can now interactively explore how a model performs on new text samples, compare predicted labels to training data, as well as reason about \textit{why} certain predictions were made by inspecting the nearest neighbors.
This iterative back-and-forth allows for the tighter integration of user inputs into model development, usage, and explanation, which in turn supports a more human-centered approach to AI that better incorporates end-user feedback, tasks, and perspectives.



{\renewcommand{\arraystretch}{1.2}
\begin{table}
    \centering
    \begin{tabular}{lll}
         &  data communication &  callbacks \\
         \hline
         DP 1 & one-way Python to JS & none \\
         DP 2 & two-way synchronized & none \\
         DP 3 & two-way synchronized & changes in data trigger callback handlers in Python kernel \\
    \end{tabular}
    \captionsetup{width=.5\columnwidth}
    \caption{Summary of features in the three design patterns.}
    \label{tab:summary}
\end{table}
}

\section{Best Practices and Design Guidelines}

So far, we have described three design patterns for how XAI systems can be embedded into Jupyter.
A summary of the design patterns is provided in Table \ref{tab:summary}.
In this section, we reflect on our experiences developing such systems and the lessons learned.
We distill these into a set of best practices and design guidelines that can be used to determine which design pattern of XAI system to develop for different usage scenarios.
We structure the following sections around four main considerations:
\textit{When} should we build and embed XAI systems into Jupyter? 
\textit{Who} are the target users of these XAI systems?
\textit{What} should be taken into account?
\textit{How}, and with what methodologies, should these XAI systems be designed, implemented, and evaluated?

\subsection{Support existing development environments}
We know from existing studies into interaction design that minimizing interruptions and context switching can enhance focus, reduce cognitive load, and improve task performance \cite{bederson2004interfaces, pike2009science, mark2008cost}.
As such, embedding XAI systems into notebooks should primarily be done when Jupyter is already the main computational environment of target users.
This is particularly useful when existing workflows are highly iterative, since repeatedly switching to a standalone XAI system can be more disruptive due to frequent task interruptions.
At the same time, the inverse is also true.
While Python libraries and computational notebooks are the dominant site of ML development today, increasingly, other tools such as TensorFlow.js \cite{tensorflow2015whitepaper} and WebLLM\footnote{https://llm.mlc.ai/} are being adopted that allow models to run directly in-browser.
In this scenario, designing and embedding XAI systems into Jupyter may be less appropriate since it would require users to relocate their workflow to entirely new languages and environments, which might increase barriers to adoption instead.

\subsection{Support existing workflows}
Another factor to consider when building embedded XAI systems is the typical model development workflow and associated user tasks.
Recent taxonomies have proposed that different tasks require different explanatory tools \cite{liao2020questioning, arya2019one}, and understanding the typical or existing user workflow is crucial for determining the appropriate places where interactive XAI widgets are needed.
For example, some widgets can help with data exploration and pre-processing, while other widgets are more appropriate for inspecting feature weights after training.
In cases where multiple XAI systems are needed to support user input at different steps of the workflow, it may even be necessary for multiple widgets to be implemented in a single library to account for varied or iterative user feedback \cite{guo2023causalvis}.

\subsection{Select design patterns based on user tasks}
We also know from prior studies that different audiences require different explanations throughout the AI life cycle \cite{dhanorkar2021needs}.
As such, the types of interactive XAI tools that are needed, the users they support, and the associated explainability tasks are all key considerations for deciding which design pattern should be adopted when building explainability tools for Jupyter environments.

For example, if an XAI system is designed for ML developers who need to refine a model iteratively, it may be more effective to use design patterns 2 and 3 to integrate explanation and inspection into the process of model development.
Similarly, if a model requires in-depth or lengthy collaboration with domain-experts or end-users without ML/data science experience, design patterns 2 and 3 may also be well-suited to help them interactively specify their requirements and domain knowledge.
In both of these cases, the automatic propagation of inputs to the Python widget means that any feedback can be immediately incorporated into the workflow, allowing developers and end-users to collaboratively discuss and iterate on ML models rapidly.
On the other hand, if user tasks are oriented around simply communicating or understanding the data or model, then the self-contained XAI systems of design pattern 1 may be sufficient.

\subsection{Design for relevant data types and libraries}
Since embedded XAI systems are meant for Python/Jupyter environments, there is also a need to account for the data types and data-processing libraries that are typically available.
In our prior work on causal inference \cite{guo2023causalvis}, for example, data analysts we collaborated with wanted the XAI tools to accept directed acyclic graph (DAG) specifications in both \verb|.json| and NetworkX\footnote{https://networkx.org/}.
This latter requirement arose because our collaborators wanted to use the CausalNex library \cite{BeaumontCausalNex2021}, which outputs NetworkX graphs, in conjunction with the interactive visualizations we provided.
Similarly, in our DataSelector example above (Section \ref{dataselector}), since the widget is developed for Hugging Face models, the training data input and the selected output of the widget are formatted as Hugging Face Dataset classes.
In both cases, while it is possible for users to manually convert data into the required data types, we can support more seamless integration of XAI systems by automatically processing the data to match available libraries and workflows.

\subsection{Incorporate computational environment throughout XAI development}
The prior sections have highlighted how, like all human-centered computing systems, interactive XAI should be implemented with a clear understanding of users and user tasks.
In visualization research, many existing methodologies are well-suited for this purpose \cite{sedlmair2012design, mccurdy2016action, syeda2020design, mckenna2014design}.
More broadly, participatory design and participatory action research approaches \cite{wadsworth1993participatory, hayes2011relationship} that emphasize designing \textit{with} users have also been proposed as effective methods for human-centered XAI \cite{ehsan2020human}.

At the same time, developing interactive XAI systems for Jupyter requires existing methodologies to be extended on account of the particularities of this computational environment.
In our prior works, we have utilized an approach that makes Jupyter part of the design process from the very beginning.
For instance, during initial prototyping, we included notebook cells as screenshots to help users visualize the intended usage context.
This was effective for eliciting requests about the data types and file formats the embedded XAI system should support.
Similarly, in the evaluation studies, we presented users with notebooks that demonstrate an end-to-end workflow with the XAI systems embedded.
This gave users a richer understanding of what the integrated workflow looks like and how the XAI systems could be added to their existing projects.
It also helped elicit feedback about how the XAI systems could facilitate collaborative discussions and analyses with other stakeholders.

\section{Open Questions} \label{open_questions}

While the examples in this paper and in our supplemental toolkit have demonstrated the possibilities of embedded XAI systems in Jupyter, as well as how they might realize integrated and collaborative ML workflows, there remain challenges to the development of these systems.
In the following sections, we discuss some of these open questions and outline potential avenues for future research.

\subsection{Data Size Limitations}

Since all communication and synchronization between the front-end interface and back-end kernel must be done in JSON format, this places a limitation on the types of information that can be transferred.
As such, while most data file types can be easily JSON serialized, it would not be possible to do the same for the models themselves.
Furthermore, there also exists a limit on the amount of data that can be transferred this way.
In our example toolkit, the XAI widgets are designed for a model fine-tuning workflow, which requires only relatively small amounts of text data.
However, other usage scenarios may include greater amounts of data or larger file types (such as image and video files) that cannot be easily transferred and displayed simultaneously.
In our prior work, we have addressed this by processing data on-demand, allowing users to interactively choose which subset of the data to serialize, communicate, and display.
This ensures that the JSON created is of a manageable size, however, it also limits the ability of users to gain an overview of the entire data set at once.
More work is thus needed to explore other solutions for the communication of large data sets, particularly in usage contexts that need also handle dynamic user interactions and interface updates.

\subsection{Alternate Design Patterns}


The design patterns included in Section \ref{designpatterns} describe the current state of embedded XAI systems in Jupyter, both in terms of existing libraries and what we have been able to achieve in our own work.
However, this is by no means an exhaustive list.
There are other approaches to implementing how JavaScript front-end interfaces can be incorporated into Python workflows.
For example, libraries such as OmniXAI \cite{wenzhuo2022omnixai}, Plotly Dash \cite{plotly} and Shapash\footnote{https://shapash.readthedocs.io} allow users to initialize an XAI system as a web app from their notebook.
This web app is hosted in a new browser window, and include interactive visualizations that help users inspect properties of their model.
While this requires users to switch between browser windows, their interactions with the front-end XAI system can still be used to communicate user-input data back to the notebook.
Future work can look at how such implementations compare to the design patterns discussed in this paper, and how they support end-user engagement in ML workflows.
More broadly, we also see potential for the synthesis of a more comprehensive taxonomy of design patterns for different types of XAI tools for computational notebooks, as well as the systematic exploration of the different usage scenarios and tasks that can be supported by each.

\subsection{Beyond Jupyter} \label{beyond_jupyterlab}

Finally, while this paper and related examples have all been developed for Jupyter, more work is needed before the design patterns discussed here can be applied to other computational environments.
Of these, Google Colab (an in-browser version of JupyterLab) is widely used by ML developers \cite{kaggle} for quick sharing and reproduction of ML workflows.
However, due to security reasons, Colab and other computational environments do not yet support the two-way communication and synchronization of data.
This thus limits the applicability of design patterns 2 and 3, which are only usable in local versions of Jupyter for now.

\section{Conclusion}

XAI tools are effective methods of explaining and reasoning about how ML models are developed and used.
However, unlike many other ML libraries and resources, most interactive XAI tools are not built for Python computational environments such as Jupyter.
In this paper, we addressed this mismatch by proposing three design patterns for how front-end XAI interfaces can be embedded into notebooks, namely: 1) One-way communication from Python to JavaScript, 2) Two-way data synchronization, and 3) Bi-directional callbacks.
We also provided an open-source toolkit, bonXAI, that provides examples of how each design pattern might be used to build interactive XAI systems that fit into a text classification model development workflow.
Taken together, these examples demonstrate how the design patterns proposed can support the better communication of explanations of ML processes, as well as methods for seamlessly incorporating end-user inputs into model development workflows.

\begin{acks}
The authors would like to thank Raj Shah for improving the InferenceExplorer embedding space. This work is supported in part by NSF IIS-1750474 and the IBM PhD Fellowship.
\end{acks}

\bibliographystyle{ACM-Reference-Format}
\bibliography{sample-base}


\begin{thebibliography}{98}


\ifx \showCODEN    \undefined \def \showCODEN     #1{\unskip}     \fi
\ifx \showDOI      \undefined \def \showDOI       #1{#1}\fi
\ifx \showISBNx    \undefined \def \showISBNx     #1{\unskip}     \fi
\ifx \showISBNxiii \undefined \def \showISBNxiii  #1{\unskip}     \fi
\ifx \showISSN     \undefined \def \showISSN      #1{\unskip}     \fi
\ifx \showLCCN     \undefined \def \showLCCN      #1{\unskip}     \fi
\ifx \shownote     \undefined \def \shownote      #1{#1}          \fi
\ifx \showarticletitle \undefined \def \showarticletitle #1{#1}   \fi
\ifx \showURL      \undefined \def \showURL       {\relax}        \fi
\providecommand\bibfield[2]{#2}
\providecommand\bibinfo[2]{#2}
\providecommand\natexlab[1]{#1}
\providecommand\showeprint[2][]{arXiv:#2}

\bibitem[A.~Leite et~al\mbox{.}(2020)]%
        {a2020neva}
\bibfield{author}{\bibinfo{person}{Roger A.~Leite}, \bibinfo{person}{Theresia Gschwandtner}, \bibinfo{person}{Silvia Miksch}, \bibinfo{person}{Erich Gstrein}, {and} \bibinfo{person}{Johannes Kuntner}.} \bibinfo{year}{2020}\natexlab{}.
\newblock \showarticletitle{Neva: Visual analytics to identify fraudulent networks}. In \bibinfo{booktitle}{\emph{Computer Graphics Forum}}, Vol.~\bibinfo{volume}{39}. Wiley Online Library, \bibinfo{pages}{344--359}.
\newblock


\bibitem[Abadi et~al\mbox{.}(2015)]%
        {tensorflow2015whitepaper}
\bibfield{author}{\bibinfo{person}{Mart\'{i}n Abadi}, \bibinfo{person}{Ashish Agarwal}, \bibinfo{person}{Paul Barham}, \bibinfo{person}{Eugene Brevdo}, \bibinfo{person}{Zhifeng Chen}, \bibinfo{person}{Craig Citro}, \bibinfo{person}{Greg~S. Corrado}, \bibinfo{person}{Andy Davis}, \bibinfo{person}{Jeffrey Dean}, \bibinfo{person}{Matthieu Devin}, \bibinfo{person}{Sanjay Ghemawat}, \bibinfo{person}{Ian Goodfellow}, \bibinfo{person}{Andrew Harp}, \bibinfo{person}{Geoffrey Irving}, \bibinfo{person}{Michael Isard}, \bibinfo{person}{Yangqing Jia}, \bibinfo{person}{Rafal Jozefowicz}, \bibinfo{person}{Lukasz Kaiser}, \bibinfo{person}{Manjunath Kudlur}, \bibinfo{person}{Josh Levenberg}, \bibinfo{person}{Dandelion Man\'{e}}, \bibinfo{person}{Rajat Monga}, \bibinfo{person}{Sherry Moore}, \bibinfo{person}{Derek Murray}, \bibinfo{person}{Chris Olah}, \bibinfo{person}{Mike Schuster}, \bibinfo{person}{Jonathon Shlens}, \bibinfo{person}{Benoit Steiner}, \bibinfo{person}{Ilya Sutskever}, \bibinfo{person}{Kunal Talwar},
  \bibinfo{person}{Paul Tucker}, \bibinfo{person}{Vincent Vanhoucke}, \bibinfo{person}{Vijay Vasudevan}, \bibinfo{person}{Fernanda Vi\'{e}gas}, \bibinfo{person}{Oriol Vinyals}, \bibinfo{person}{Pete Warden}, \bibinfo{person}{Martin Wattenberg}, \bibinfo{person}{Martin Wicke}, \bibinfo{person}{Yuan Yu}, {and} \bibinfo{person}{Xiaoqiang Zheng}.} \bibinfo{year}{2015}\natexlab{}.
\newblock \bibinfo{title}{{TensorFlow}: Large-Scale Machine Learning on Heterogeneous Systems}.
\newblock
\newblock
\urldef\tempurl%
\url{https://www.tensorflow.org/}
\showURL{%
\tempurl}
\newblock
\shownote{Software available from tensorflow.org}.


\bibitem[Ackerman(2000)]%
        {ackerman2000intellectual}
\bibfield{author}{\bibinfo{person}{Mark~S Ackerman}.} \bibinfo{year}{2000}\natexlab{}.
\newblock \showarticletitle{The intellectual challenge of CSCW: the gap between social requirements and technical feasibility}.
\newblock \bibinfo{journal}{\emph{Human--Computer Interaction}} \bibinfo{volume}{15}, \bibinfo{number}{2-3} (\bibinfo{year}{2000}), \bibinfo{pages}{179--203}.
\newblock


\bibitem[Adadi and Berrada(2018)]%
        {adadi2018peeking}
\bibfield{author}{\bibinfo{person}{Amina Adadi} {and} \bibinfo{person}{Mohammed Berrada}.} \bibinfo{year}{2018}\natexlab{}.
\newblock \showarticletitle{Peeking inside the black-box: a survey on explainable artificial intelligence (XAI)}.
\newblock \bibinfo{journal}{\emph{IEEE access}}  \bibinfo{volume}{6} (\bibinfo{year}{2018}), \bibinfo{pages}{52138--52160}.
\newblock


\bibitem[Ahn et~al\mbox{.}(2022)]%
        {ahn2022tribe}
\bibfield{author}{\bibinfo{person}{Yongsu Ahn}, \bibinfo{person}{Muheng Yan}, \bibinfo{person}{Yu-Ru Lin}, \bibinfo{person}{Wen-Ting Chung}, {and} \bibinfo{person}{Rebecca Hwa}.} \bibinfo{year}{2022}\natexlab{}.
\newblock \showarticletitle{Tribe or not? Critical inspection of group differences using TribalGram}.
\newblock \bibinfo{journal}{\emph{ACM Transactions on Interactive Intelligent Systems (TiiS)}} \bibinfo{volume}{12}, \bibinfo{number}{1} (\bibinfo{year}{2022}), \bibinfo{pages}{1--34}.
\newblock


\bibitem[Alammar(2021)]%
        {alammar2021ecco}
\bibfield{author}{\bibinfo{person}{J Alammar}.} \bibinfo{year}{2021}\natexlab{}.
\newblock \showarticletitle{Ecco: An Open Source Library for the Explainability of Transformer Language Models}. In \bibinfo{booktitle}{\emph{Proceedings of the 59th Annual Meeting of the Association for Computational Linguistics and the 11th International Joint Conference on Natural Language Processing: System Demonstrations}}. \bibinfo{publisher}{Association for Computational Linguistics}, \bibinfo{pages}{249--257}.
\newblock


\bibitem[Arrieta et~al\mbox{.}(2020)]%
        {arrieta2020explainable}
\bibfield{author}{\bibinfo{person}{Alejandro~Barredo Arrieta}, \bibinfo{person}{Natalia D{\'\i}az-Rodr{\'\i}guez}, \bibinfo{person}{Javier Del~Ser}, \bibinfo{person}{Adrien Bennetot}, \bibinfo{person}{Siham Tabik}, \bibinfo{person}{Alberto Barbado}, \bibinfo{person}{Salvador Garc{\'\i}a}, \bibinfo{person}{Sergio Gil-L{\'o}pez}, \bibinfo{person}{Daniel Molina}, \bibinfo{person}{Richard Benjamins}, {et~al\mbox{.}}} \bibinfo{year}{2020}\natexlab{}.
\newblock \showarticletitle{Explainable Artificial Intelligence (XAI): Concepts, taxonomies, opportunities and challenges toward responsible AI}.
\newblock \bibinfo{journal}{\emph{Information fusion}}  \bibinfo{volume}{58} (\bibinfo{year}{2020}), \bibinfo{pages}{82--115}.
\newblock


\bibitem[Arya et~al\mbox{.}(2019)]%
        {arya2019one}
\bibfield{author}{\bibinfo{person}{Vijay Arya}, \bibinfo{person}{Rachel~KE Bellamy}, \bibinfo{person}{Pin-Yu Chen}, \bibinfo{person}{Amit Dhurandhar}, \bibinfo{person}{Michael Hind}, \bibinfo{person}{Samuel~C Hoffman}, \bibinfo{person}{Stephanie Houde}, \bibinfo{person}{Q~Vera Liao}, \bibinfo{person}{Ronny Luss}, \bibinfo{person}{Aleksandra Mojsilovi{\'c}}, {et~al\mbox{.}}} \bibinfo{year}{2019}\natexlab{}.
\newblock \showarticletitle{One explanation does not fit all: A toolkit and taxonomy of ai explainability techniques}.
\newblock \bibinfo{journal}{\emph{arXiv preprint arXiv:1909.03012}} (\bibinfo{year}{2019}).
\newblock


\bibitem[Beaumont et~al\mbox{.}(2021)]%
        {BeaumontCausalNex2021}
\bibfield{author}{\bibinfo{person}{Paul Beaumont}, \bibinfo{person}{Ben Horsburgh}, \bibinfo{person}{Philip Pilgerstorfer}, \bibinfo{person}{Angel Droth}, \bibinfo{person}{Richard Oentaryo}, \bibinfo{person}{Steven Ler}, \bibinfo{person}{Hiep Nguyen}, \bibinfo{person}{Gabriel~Azevedo Ferreira}, \bibinfo{person}{Zain Patel}, {and} \bibinfo{person}{Wesley Leong}.} \bibinfo{year}{2021}\natexlab{}.
\newblock \bibinfo{booktitle}{\emph{{CausalNex}}}.
\newblock
\urldef\tempurl%
\url{https://github.com/quantumblacklabs/causalnex}
\showURL{%
\tempurl}


\bibitem[Bederson(2004)]%
        {bederson2004interfaces}
\bibfield{author}{\bibinfo{person}{Benjamin~B Bederson}.} \bibinfo{year}{2004}\natexlab{}.
\newblock \showarticletitle{Interfaces for staying in the flow}.
\newblock \bibinfo{journal}{\emph{Ubiquity}} \bibinfo{volume}{5}, \bibinfo{number}{27} (\bibinfo{year}{2004}), \bibinfo{pages}{1}.
\newblock


\bibitem[Bertrand et~al\mbox{.}(2023)]%
        {bertrand2023selective}
\bibfield{author}{\bibinfo{person}{Astrid Bertrand}, \bibinfo{person}{Tiphaine Viard}, \bibinfo{person}{Rafik Belloum}, \bibinfo{person}{James~R Eagan}, {and} \bibinfo{person}{Winston Maxwell}.} \bibinfo{year}{2023}\natexlab{}.
\newblock \showarticletitle{On Selective, Mutable and Dialogic XAI: a Review of What Users Say about Different Types of Interactive Explanations}. In \bibinfo{booktitle}{\emph{Proceedings of the 2023 CHI Conference on Human Factors in Computing Systems}}. \bibinfo{pages}{1--21}.
\newblock


\bibitem[Bhatt et~al\mbox{.}(2020)]%
        {bhatt2020explainable}
\bibfield{author}{\bibinfo{person}{Umang Bhatt}, \bibinfo{person}{Alice Xiang}, \bibinfo{person}{Shubham Sharma}, \bibinfo{person}{Adrian Weller}, \bibinfo{person}{Ankur Taly}, \bibinfo{person}{Yunhan Jia}, \bibinfo{person}{Joydeep Ghosh}, \bibinfo{person}{Ruchir Puri}, \bibinfo{person}{Jos{\'e}~MF Moura}, {and} \bibinfo{person}{Peter Eckersley}.} \bibinfo{year}{2020}\natexlab{}.
\newblock \showarticletitle{Explainable machine learning in deployment}. In \bibinfo{booktitle}{\emph{Proceedings of the 2020 conference on fairness, accountability, and transparency}}. \bibinfo{pages}{648--657}.
\newblock


\bibitem[Cai et~al\mbox{.}(2019a)]%
        {cai2019human}
\bibfield{author}{\bibinfo{person}{Carrie~J Cai}, \bibinfo{person}{Emily Reif}, \bibinfo{person}{Narayan Hegde}, \bibinfo{person}{Jason Hipp}, \bibinfo{person}{Been Kim}, \bibinfo{person}{Daniel Smilkov}, \bibinfo{person}{Martin Wattenberg}, \bibinfo{person}{Fernanda Viegas}, \bibinfo{person}{Greg~S Corrado}, \bibinfo{person}{Martin~C Stumpe}, {et~al\mbox{.}}} \bibinfo{year}{2019}\natexlab{a}.
\newblock \showarticletitle{Human-centered tools for coping with imperfect algorithms during medical decision-making}. In \bibinfo{booktitle}{\emph{Proceedings of the 2019 chi conference on human factors in computing systems}}. \bibinfo{pages}{1--14}.
\newblock


\bibitem[Cai et~al\mbox{.}(2019b)]%
        {cai2019hello}
\bibfield{author}{\bibinfo{person}{Carrie~J Cai}, \bibinfo{person}{Samantha Winter}, \bibinfo{person}{David Steiner}, \bibinfo{person}{Lauren Wilcox}, {and} \bibinfo{person}{Michael Terry}.} \bibinfo{year}{2019}\natexlab{b}.
\newblock \showarticletitle{" Hello AI": uncovering the onboarding needs of medical practitioners for human-AI collaborative decision-making}.
\newblock \bibinfo{journal}{\emph{Proceedings of the ACM on Human-computer Interaction}} \bibinfo{volume}{3}, \bibinfo{number}{CSCW} (\bibinfo{year}{2019}), \bibinfo{pages}{1--24}.
\newblock


\bibitem[Chollet et~al\mbox{.}(2015)]%
        {chollet2015keras}
\bibfield{author}{\bibinfo{person}{Fran\c{c}ois Chollet} {et~al\mbox{.}}} \bibinfo{year}{2015}\natexlab{}.
\newblock \bibinfo{title}{Keras}.
\newblock \bibinfo{howpublished}{\url{https://keras.io}}.
\newblock


\bibitem[Christiano et~al\mbox{.}(2017)]%
        {christiano2017deep}
\bibfield{author}{\bibinfo{person}{Paul~F Christiano}, \bibinfo{person}{Jan Leike}, \bibinfo{person}{Tom Brown}, \bibinfo{person}{Miljan Martic}, \bibinfo{person}{Shane Legg}, {and} \bibinfo{person}{Dario Amodei}.} \bibinfo{year}{2017}\natexlab{}.
\newblock \showarticletitle{Deep reinforcement learning from human preferences}.
\newblock \bibinfo{journal}{\emph{Advances in neural information processing systems}}  \bibinfo{volume}{30} (\bibinfo{year}{2017}).
\newblock


\bibitem[Corbett et~al\mbox{.}(1997)]%
        {corbett1997intelligent}
\bibfield{author}{\bibinfo{person}{Albert~T Corbett}, \bibinfo{person}{Kenneth~R Koedinger}, {and} \bibinfo{person}{John~R Anderson}.} \bibinfo{year}{1997}\natexlab{}.
\newblock \showarticletitle{Intelligent tutoring systems}.
\newblock In \bibinfo{booktitle}{\emph{Handbook of human-computer interaction}}. \bibinfo{publisher}{Elsevier}, \bibinfo{pages}{849--874}.
\newblock


\bibitem[Dhanorkar et~al\mbox{.}(2021)]%
        {dhanorkar2021needs}
\bibfield{author}{\bibinfo{person}{Shipi Dhanorkar}, \bibinfo{person}{Christine~T Wolf}, \bibinfo{person}{Kun Qian}, \bibinfo{person}{Anbang Xu}, \bibinfo{person}{Lucian Popa}, {and} \bibinfo{person}{Yunyao Li}.} \bibinfo{year}{2021}\natexlab{}.
\newblock \showarticletitle{Who needs to know what, when?: Broadening the Explainable AI (XAI) Design Space by Looking at Explanations Across the AI Lifecycle}. In \bibinfo{booktitle}{\emph{Designing Interactive Systems Conference 2021}}. \bibinfo{pages}{1591--1602}.
\newblock


\bibitem[Doshi-Velez and Kim(2017)]%
        {doshi2017towards}
\bibfield{author}{\bibinfo{person}{Finale Doshi-Velez} {and} \bibinfo{person}{Been Kim}.} \bibinfo{year}{2017}\natexlab{}.
\newblock \showarticletitle{Towards a rigorous science of interpretable machine learning}.
\newblock \bibinfo{journal}{\emph{arXiv preprint arXiv:1702.08608}} (\bibinfo{year}{2017}).
\newblock


\bibitem[Ehsan et~al\mbox{.}(2021a)]%
        {ehsan2021expanding}
\bibfield{author}{\bibinfo{person}{Upol Ehsan}, \bibinfo{person}{Q~Vera Liao}, \bibinfo{person}{Michael Muller}, \bibinfo{person}{Mark~O Riedl}, {and} \bibinfo{person}{Justin~D Weisz}.} \bibinfo{year}{2021}\natexlab{a}.
\newblock \showarticletitle{Expanding explainability: Towards social transparency in ai systems}. In \bibinfo{booktitle}{\emph{Proceedings of the 2021 CHI Conference on Human Factors in Computing Systems}}. \bibinfo{pages}{1--19}.
\newblock


\bibitem[Ehsan et~al\mbox{.}(2021b)]%
        {ehsan2021explainable}
\bibfield{author}{\bibinfo{person}{Upol Ehsan}, \bibinfo{person}{Samir Passi}, \bibinfo{person}{Q~Vera Liao}, \bibinfo{person}{Larry Chan}, \bibinfo{person}{I Lee}, \bibinfo{person}{Michael Muller}, \bibinfo{person}{Mark~O Riedl}, {et~al\mbox{.}}} \bibinfo{year}{2021}\natexlab{b}.
\newblock \showarticletitle{The who in explainable ai: How ai background shapes perceptions of ai explanations}.
\newblock \bibinfo{journal}{\emph{arXiv preprint arXiv:2107.13509}} (\bibinfo{year}{2021}).
\newblock


\bibitem[Ehsan and Riedl(2020)]%
        {ehsan2020human}
\bibfield{author}{\bibinfo{person}{Upol Ehsan} {and} \bibinfo{person}{Mark~O Riedl}.} \bibinfo{year}{2020}\natexlab{}.
\newblock \showarticletitle{Human-centered explainable ai: Towards a reflective sociotechnical approach}. In \bibinfo{booktitle}{\emph{HCI International 2020-Late Breaking Papers: Multimodality and Intelligence: 22nd HCI International Conference, HCII 2020, Copenhagen, Denmark, July 19--24, 2020, Proceedings 22}}. Springer, \bibinfo{pages}{449--466}.
\newblock


\bibitem[Ehsan et~al\mbox{.}(2023)]%
        {ehsan2023charting}
\bibfield{author}{\bibinfo{person}{Upol Ehsan}, \bibinfo{person}{Koustuv Saha}, \bibinfo{person}{Munmun De~Choudhury}, {and} \bibinfo{person}{Mark~O Riedl}.} \bibinfo{year}{2023}\natexlab{}.
\newblock \showarticletitle{Charting the Sociotechnical Gap in Explainable AI: A Framework to Address the Gap in XAI}.
\newblock \bibinfo{journal}{\emph{Proceedings of the ACM on Human-Computer Interaction}} \bibinfo{volume}{7}, \bibinfo{number}{CSCW1} (\bibinfo{year}{2023}), \bibinfo{pages}{1--32}.
\newblock


\bibitem[Ehsan et~al\mbox{.}(2022a)]%
        {ehsan2022algorithmic}
\bibfield{author}{\bibinfo{person}{Upol Ehsan}, \bibinfo{person}{Ranjit Singh}, \bibinfo{person}{Jacob Metcalf}, {and} \bibinfo{person}{Mark Riedl}.} \bibinfo{year}{2022}\natexlab{a}.
\newblock \showarticletitle{The algorithmic imprint}. In \bibinfo{booktitle}{\emph{Proceedings of the 2022 ACM Conference on Fairness, Accountability, and Transparency}}. \bibinfo{pages}{1305--1317}.
\newblock


\bibitem[Ehsan et~al\mbox{.}(2021c)]%
        {ehsan2021operationalizing}
\bibfield{author}{\bibinfo{person}{Upol Ehsan}, \bibinfo{person}{Philipp Wintersberger}, \bibinfo{person}{Q~Vera Liao}, \bibinfo{person}{Martina Mara}, \bibinfo{person}{Marc Streit}, \bibinfo{person}{Sandra Wachter}, \bibinfo{person}{Andreas Riener}, {and} \bibinfo{person}{Mark~O Riedl}.} \bibinfo{year}{2021}\natexlab{c}.
\newblock \showarticletitle{Operationalizing human-centered perspectives in explainable AI}. In \bibinfo{booktitle}{\emph{Extended Abstracts of the 2021 CHI Conference on Human Factors in Computing Systems}}. \bibinfo{pages}{1--6}.
\newblock


\bibitem[Ehsan et~al\mbox{.}(2022b)]%
        {ehsan2022human}
\bibfield{author}{\bibinfo{person}{Upol Ehsan}, \bibinfo{person}{Philipp Wintersberger}, \bibinfo{person}{Q~Vera Liao}, \bibinfo{person}{Elizabeth~Anne Watkins}, \bibinfo{person}{Carina Manger}, \bibinfo{person}{Hal Daum{\'e}~III}, \bibinfo{person}{Andreas Riener}, {and} \bibinfo{person}{Mark~O Riedl}.} \bibinfo{year}{2022}\natexlab{b}.
\newblock \showarticletitle{Human-Centered Explainable AI (HCXAI): beyond opening the black-box of AI}. In \bibinfo{booktitle}{\emph{CHI conference on human factors in computing systems extended abstracts}}. \bibinfo{pages}{1--7}.
\newblock


\bibitem[eli5 community(2021)]%
        {ELI5}
\bibfield{author}{\bibinfo{person}{eli5 community}.} \bibinfo{year}{2021}\natexlab{}.
\newblock \bibinfo{title}{Welcome to ELI5’s documentation!}
\newblock
\newblock
\urldef\tempurl%
\url{https://eli5.readthedocs.io/en/latest/}
\showURL{%
\tempurl}
\newblock
\shownote{Accessed Feb 20, 2024}.


\bibitem[Face({[n.\,d.]})]%
        {HuggingFacetextclassification}
\bibfield{author}{\bibinfo{person}{Hugging Face}.} \bibinfo{year}{[n.\,d.]}\natexlab{}.
\newblock \bibinfo{title}{Text classification}.
\newblock
\newblock
\urldef\tempurl%
\url{https://huggingface.co/docs/transformers/en/tasks/sequence_classification}
\showURL{%
\tempurl}
\newblock
\shownote{Accessed Feb 20, 2024}.


\bibitem[Fiok et~al\mbox{.}(2022)]%
        {fiok2022explainable}
\bibfield{author}{\bibinfo{person}{Krzysztof Fiok}, \bibinfo{person}{Farzad~V Farahani}, \bibinfo{person}{Waldemar Karwowski}, {and} \bibinfo{person}{Tareq Ahram}.} \bibinfo{year}{2022}\natexlab{}.
\newblock \showarticletitle{Explainable artificial intelligence for education and training}.
\newblock \bibinfo{journal}{\emph{The Journal of Defense Modeling and Simulation}} \bibinfo{volume}{19}, \bibinfo{number}{2} (\bibinfo{year}{2022}), \bibinfo{pages}{133--144}.
\newblock


\bibitem[Gildenblat and contributors(2021)]%
        {jacobgilpytorchcam}
\bibfield{author}{\bibinfo{person}{Jacob Gildenblat} {and} \bibinfo{person}{contributors}.} \bibinfo{year}{2021}\natexlab{}.
\newblock \bibinfo{title}{PyTorch library for CAM methods}.
\newblock \bibinfo{howpublished}{\url{https://github.com/jacobgil/pytorch-grad-cam}}.
\newblock


\bibitem[Guo et~al\mbox{.}(2021)]%
        {guo2021vaine}
\bibfield{author}{\bibinfo{person}{Grace Guo}, \bibinfo{person}{Maria Glenski}, \bibinfo{person}{ZhuanYi Shaw}, \bibinfo{person}{Emily Saldanha}, \bibinfo{person}{Alex Endert}, \bibinfo{person}{Svitlana Volkova}, {and} \bibinfo{person}{Dustin Arendt}.} \bibinfo{year}{2021}\natexlab{}.
\newblock \showarticletitle{Vaine: Visualization and ai for natural experiments}. In \bibinfo{booktitle}{\emph{2021 IEEE Visualization Conference (VIS)}}. IEEE, \bibinfo{pages}{21--25}.
\newblock


\bibitem[Guo et~al\mbox{.}(2023)]%
        {guo2023causalvis}
\bibfield{author}{\bibinfo{person}{Grace Guo}, \bibinfo{person}{Ehud Karavani}, \bibinfo{person}{Alex Endert}, {and} \bibinfo{person}{Bum~Chul Kwon}.} \bibinfo{year}{2023}\natexlab{}.
\newblock \showarticletitle{Causalvis: Visualizations for Causal Inference}. In \bibinfo{booktitle}{\emph{Proceedings of the 2023 CHI Conference on Human Factors in Computing Systems}}. \bibinfo{pages}{1--20}.
\newblock


\bibitem[Hagerty and Rubinov(2019)]%
        {hagerty2019global}
\bibfield{author}{\bibinfo{person}{Alexa Hagerty} {and} \bibinfo{person}{Igor Rubinov}.} \bibinfo{year}{2019}\natexlab{}.
\newblock \showarticletitle{Global AI ethics: a review of the social impacts and ethical implications of artificial intelligence}.
\newblock \bibinfo{journal}{\emph{arXiv preprint arXiv:1907.07892}} (\bibinfo{year}{2019}).
\newblock


\bibitem[Hayes(2011)]%
        {hayes2011relationship}
\bibfield{author}{\bibinfo{person}{Gillian~R Hayes}.} \bibinfo{year}{2011}\natexlab{}.
\newblock \showarticletitle{The relationship of action research to human-computer interaction}.
\newblock \bibinfo{journal}{\emph{ACM Transactions on Computer-Human Interaction (TOCHI)}} \bibinfo{volume}{18}, \bibinfo{number}{3} (\bibinfo{year}{2011}), \bibinfo{pages}{1--20}.
\newblock


\bibitem[Hong et~al\mbox{.}(2020)]%
        {hong2020human}
\bibfield{author}{\bibinfo{person}{Sungsoo~Ray Hong}, \bibinfo{person}{Jessica Hullman}, {and} \bibinfo{person}{Enrico Bertini}.} \bibinfo{year}{2020}\natexlab{}.
\newblock \showarticletitle{Human factors in model interpretability: Industry practices, challenges, and needs}.
\newblock \bibinfo{journal}{\emph{Proceedings of the ACM on Human-Computer Interaction}} \bibinfo{volume}{4}, \bibinfo{number}{CSCW1} (\bibinfo{year}{2020}), \bibinfo{pages}{1--26}.
\newblock


\bibitem[Hunter(2007)]%
        {Hunter2007}
\bibfield{author}{\bibinfo{person}{J.~D. Hunter}.} \bibinfo{year}{2007}\natexlab{}.
\newblock \showarticletitle{Matplotlib: A 2D graphics environment}.
\newblock \bibinfo{journal}{\emph{Computing in Science \& Engineering}} \bibinfo{volume}{9}, \bibinfo{number}{3} (\bibinfo{year}{2007}), \bibinfo{pages}{90--95}.
\newblock
\urldef\tempurl%
\url{https://doi.org/10.1109/MCSE.2007.55}
\showDOI{\tempurl}


\bibitem[Inc.(2015)]%
        {plotly}
\bibfield{author}{\bibinfo{person}{Plotly~Technologies Inc.}} \bibinfo{year}{2015}\natexlab{}.
\newblock \bibinfo{booktitle}{\emph{Collaborative data science}}.
\newblock Montreal, QC.
\newblock
\urldef\tempurl%
\url{https://plot.ly}
\showURL{%
\tempurl}


\bibitem[Kaggle(2022)]%
        {kaggle}
\bibfield{author}{\bibinfo{person}{Kaggle}.} \bibinfo{year}{2022}\natexlab{}.
\newblock \bibinfo{title}{State of Data Science and Machine Learning 2022}.
\newblock
\newblock
\urldef\tempurl%
\url{https://www.kaggle.com/kaggle-survey-2022}
\showURL{%
\tempurl}
\newblock
\shownote{Accessed Feb 20, 2024}.


\bibitem[Kang et~al\mbox{.}(2021)]%
        {kang2021toonnote}
\bibfield{author}{\bibinfo{person}{DaYe Kang}, \bibinfo{person}{Tony Ho}, \bibinfo{person}{Nicolai Marquardt}, \bibinfo{person}{Bilge Mutlu}, {and} \bibinfo{person}{Andrea Bianchi}.} \bibinfo{year}{2021}\natexlab{}.
\newblock \showarticletitle{Toonnote: Improving communication in computational notebooks using interactive data comics}. In \bibinfo{booktitle}{\emph{Proceedings of the 2021 CHI Conference on Human Factors in Computing Systems}}. \bibinfo{pages}{1--14}.
\newblock


\bibitem[Kery et~al\mbox{.}(2018)]%
        {kery2018story}
\bibfield{author}{\bibinfo{person}{Mary~Beth Kery}, \bibinfo{person}{Marissa Radensky}, \bibinfo{person}{Mahima Arya}, \bibinfo{person}{Bonnie~E John}, {and} \bibinfo{person}{Brad~A Myers}.} \bibinfo{year}{2018}\natexlab{}.
\newblock \showarticletitle{The story in the notebook: Exploratory data science using a literate programming tool}. In \bibinfo{booktitle}{\emph{Proceedings of the 2018 CHI conference on human factors in computing systems}}. \bibinfo{pages}{1--11}.
\newblock


\bibitem[Kery et~al\mbox{.}(2020)]%
        {kery2020mage}
\bibfield{author}{\bibinfo{person}{Mary~Beth Kery}, \bibinfo{person}{Donghao Ren}, \bibinfo{person}{Fred Hohman}, \bibinfo{person}{Dominik Moritz}, \bibinfo{person}{Kanit Wongsuphasawat}, {and} \bibinfo{person}{Kayur Patel}.} \bibinfo{year}{2020}\natexlab{}.
\newblock \showarticletitle{mage: Fluid moves between code and graphical work in computational notebooks}. In \bibinfo{booktitle}{\emph{Proceedings of the 33rd Annual ACM Symposium on User Interface Software and Technology}}. \bibinfo{pages}{140--151}.
\newblock


\bibitem[Khosravi et~al\mbox{.}(2022)]%
        {khosravi2022explainable}
\bibfield{author}{\bibinfo{person}{Hassan Khosravi}, \bibinfo{person}{Simon~Buckingham Shum}, \bibinfo{person}{Guanliang Chen}, \bibinfo{person}{Cristina Conati}, \bibinfo{person}{Yi-Shan Tsai}, \bibinfo{person}{Judy Kay}, \bibinfo{person}{Simon Knight}, \bibinfo{person}{Roberto Martinez-Maldonado}, \bibinfo{person}{Shazia Sadiq}, {and} \bibinfo{person}{Dragan Ga{\v{s}}evi{\'c}}.} \bibinfo{year}{2022}\natexlab{}.
\newblock \showarticletitle{Explainable artificial intelligence in education}.
\newblock \bibinfo{journal}{\emph{Computers and Education: Artificial Intelligence}}  \bibinfo{volume}{3} (\bibinfo{year}{2022}), \bibinfo{pages}{100074}.
\newblock


\bibitem[Kim et~al\mbox{.}(2023)]%
        {kim2023help}
\bibfield{author}{\bibinfo{person}{Sunnie~SY Kim}, \bibinfo{person}{Elizabeth~Anne Watkins}, \bibinfo{person}{Olga Russakovsky}, \bibinfo{person}{Ruth Fong}, {and} \bibinfo{person}{Andr{\'e}s Monroy-Hern{\'a}ndez}.} \bibinfo{year}{2023}\natexlab{}.
\newblock \showarticletitle{" Help Me Help the AI": Understanding How Explainability Can Support Human-AI Interaction}. In \bibinfo{booktitle}{\emph{Proceedings of the 2023 CHI Conference on Human Factors in Computing Systems}}. \bibinfo{pages}{1--17}.
\newblock


\bibitem[Kluyver et~al\mbox{.}(2016)]%
        {kluyver2016jupyter}
\bibfield{author}{\bibinfo{person}{Thomas Kluyver}, \bibinfo{person}{Benjamin Ragan-Kelley}, \bibinfo{person}{Fernando P{\'e}rez}, \bibinfo{person}{Brian~E Granger}, \bibinfo{person}{Matthias Bussonnier}, \bibinfo{person}{Jonathan Frederic}, \bibinfo{person}{Kyle Kelley}, \bibinfo{person}{Jessica~B Hamrick}, \bibinfo{person}{Jason Grout}, \bibinfo{person}{Sylvain Corlay}, {et~al\mbox{.}}} \bibinfo{year}{2016}\natexlab{}.
\newblock \showarticletitle{Jupyter Notebooks-a publishing format for reproducible computational workflows.}
\newblock \bibinfo{journal}{\emph{Elpub}}  \bibinfo{volume}{2016} (\bibinfo{year}{2016}), \bibinfo{pages}{87--90}.
\newblock


\bibitem[Knox and Stone(2008)]%
        {knox2008tamer}
\bibfield{author}{\bibinfo{person}{W~Bradley Knox} {and} \bibinfo{person}{Peter Stone}.} \bibinfo{year}{2008}\natexlab{}.
\newblock \showarticletitle{Tamer: Training an agent manually via evaluative reinforcement}. In \bibinfo{booktitle}{\emph{2008 7th IEEE international conference on development and learning}}. IEEE, \bibinfo{pages}{292--297}.
\newblock


\bibitem[Knuth(1984)]%
        {knuth1984literate}
\bibfield{author}{\bibinfo{person}{Donald~Ervin Knuth}.} \bibinfo{year}{1984}\natexlab{}.
\newblock \showarticletitle{Literate programming}.
\newblock \bibinfo{journal}{\emph{The computer journal}} \bibinfo{volume}{27}, \bibinfo{number}{2} (\bibinfo{year}{1984}), \bibinfo{pages}{97--111}.
\newblock


\bibitem[Kokhlikyan et~al\mbox{.}(2020)]%
        {kokhlikyan2020captum}
\bibfield{author}{\bibinfo{person}{Narine Kokhlikyan}, \bibinfo{person}{Vivek Miglani}, \bibinfo{person}{Miguel Martin}, \bibinfo{person}{Edward Wang}, \bibinfo{person}{Bilal Alsallakh}, \bibinfo{person}{Jonathan Reynolds}, \bibinfo{person}{Alexander Melnikov}, \bibinfo{person}{Natalia Kliushkina}, \bibinfo{person}{Carlos Araya}, \bibinfo{person}{Siqi Yan}, {and} \bibinfo{person}{Orion Reblitz-Richardson}.} \bibinfo{year}{2020}\natexlab{}.
\newblock \bibinfo{title}{Captum: A unified and generic model interpretability library for PyTorch}.
\newblock
\newblock
\showeprint[arxiv]{2009.07896}~[cs.LG]


\bibitem[Kouvela et~al\mbox{.}(2020)]%
        {kouvela2020bot}
\bibfield{author}{\bibinfo{person}{Maria Kouvela}, \bibinfo{person}{Ilias Dimitriadis}, {and} \bibinfo{person}{Athena Vakali}.} \bibinfo{year}{2020}\natexlab{}.
\newblock \showarticletitle{Bot-Detective: An explainable Twitter bot detection service with crowdsourcing functionalities}. In \bibinfo{booktitle}{\emph{Proceedings of the 12th International Conference on Management of Digital EcoSystems}}. \bibinfo{pages}{55--63}.
\newblock


\bibitem[Krishnan et~al\mbox{.}(2016a)]%
        {krishnan2016towards}
\bibfield{author}{\bibinfo{person}{Sanjay Krishnan}, \bibinfo{person}{Daniel Haas}, \bibinfo{person}{Michael~J Franklin}, {and} \bibinfo{person}{Eugene Wu}.} \bibinfo{year}{2016}\natexlab{a}.
\newblock \showarticletitle{Towards reliable interactive data cleaning: A user survey and recommendations}. In \bibinfo{booktitle}{\emph{Proceedings of the Workshop on Human-In-the-Loop Data Analytics}}. \bibinfo{pages}{1--5}.
\newblock


\bibitem[Krishnan et~al\mbox{.}(2016b)]%
        {krishnan2016activeclean}
\bibfield{author}{\bibinfo{person}{Sanjay Krishnan}, \bibinfo{person}{Jiannan Wang}, \bibinfo{person}{Eugene Wu}, \bibinfo{person}{Michael~J Franklin}, {and} \bibinfo{person}{Ken Goldberg}.} \bibinfo{year}{2016}\natexlab{b}.
\newblock \showarticletitle{Activeclean: Interactive data cleaning for statistical modeling}.
\newblock \bibinfo{journal}{\emph{Proceedings of the VLDB Endowment}} \bibinfo{volume}{9}, \bibinfo{number}{12} (\bibinfo{year}{2016}), \bibinfo{pages}{948--959}.
\newblock


\bibitem[Lambert et~al\mbox{.}(2022)]%
        {lambert2022illustrating}
\bibfield{author}{\bibinfo{person}{Nathan Lambert}, \bibinfo{person}{Louis Castricato}, \bibinfo{person}{Leandro von Werra}, {and} \bibinfo{person}{Alex Havrilla}.} \bibinfo{year}{2022}\natexlab{}.
\newblock \showarticletitle{Illustrating Reinforcement Learning from Human Feedback (RLHF)}.
\newblock \bibinfo{journal}{\emph{Hugging Face Blog}} (\bibinfo{year}{2022}).
\newblock
\newblock
\shownote{https://huggingface.co/blog/rlhf}.


\bibitem[Langer et~al\mbox{.}(2021)]%
        {langer2021we}
\bibfield{author}{\bibinfo{person}{Markus Langer}, \bibinfo{person}{Daniel Oster}, \bibinfo{person}{Timo Speith}, \bibinfo{person}{Holger Hermanns}, \bibinfo{person}{Lena K{\"a}stner}, \bibinfo{person}{Eva Schmidt}, \bibinfo{person}{Andreas Sesing}, {and} \bibinfo{person}{Kevin Baum}.} \bibinfo{year}{2021}\natexlab{}.
\newblock \showarticletitle{What do we want from Explainable Artificial Intelligence (XAI)?--A stakeholder perspective on XAI and a conceptual model guiding interdisciplinary XAI research}.
\newblock \bibinfo{journal}{\emph{Artificial Intelligence}}  \bibinfo{volume}{296} (\bibinfo{year}{2021}), \bibinfo{pages}{103473}.
\newblock


\bibitem[Leite et~al\mbox{.}(2017)]%
        {leite2017eva}
\bibfield{author}{\bibinfo{person}{Roger~A Leite}, \bibinfo{person}{Theresia Gschwandtner}, \bibinfo{person}{Silvia Miksch}, \bibinfo{person}{Simone Kriglstein}, \bibinfo{person}{Margit Pohl}, \bibinfo{person}{Erich Gstrein}, {and} \bibinfo{person}{Johannes Kuntner}.} \bibinfo{year}{2017}\natexlab{}.
\newblock \showarticletitle{Eva: Visual analytics to identify fraudulent events}.
\newblock \bibinfo{journal}{\emph{IEEE transactions on visualization and computer graphics}} \bibinfo{volume}{24}, \bibinfo{number}{1} (\bibinfo{year}{2017}), \bibinfo{pages}{330--339}.
\newblock


\bibitem[Li et~al\mbox{.}(2023)]%
        {li2023notable}
\bibfield{author}{\bibinfo{person}{Haotian Li}, \bibinfo{person}{Lu Ying}, \bibinfo{person}{Haidong Zhang}, \bibinfo{person}{Yingcai Wu}, \bibinfo{person}{Huamin Qu}, {and} \bibinfo{person}{Yun Wang}.} \bibinfo{year}{2023}\natexlab{}.
\newblock \showarticletitle{Notable: On-the-fly Assistant for Data Storytelling in Computational Notebooks}. In \bibinfo{booktitle}{\emph{Proceedings of the 2023 CHI Conference on Human Factors in Computing Systems}}. \bibinfo{pages}{1--16}.
\newblock


\bibitem[Liao et~al\mbox{.}(2020)]%
        {liao2020questioning}
\bibfield{author}{\bibinfo{person}{Q~Vera Liao}, \bibinfo{person}{Daniel Gruen}, {and} \bibinfo{person}{Sarah Miller}.} \bibinfo{year}{2020}\natexlab{}.
\newblock \showarticletitle{Questioning the AI: informing design practices for explainable AI user experiences}. In \bibinfo{booktitle}{\emph{Proceedings of the 2020 CHI conference on human factors in computing systems}}. \bibinfo{pages}{1--15}.
\newblock


\bibitem[Liao and Varshney(2021)]%
        {liao2021human}
\bibfield{author}{\bibinfo{person}{Q~Vera Liao} {and} \bibinfo{person}{Kush~R Varshney}.} \bibinfo{year}{2021}\natexlab{}.
\newblock \showarticletitle{Human-centered explainable ai (xai): From algorithms to user experiences}.
\newblock \bibinfo{journal}{\emph{arXiv preprint arXiv:2110.10790}} (\bibinfo{year}{2021}).
\newblock


\bibitem[Lundberg and Lee(2017)]%
        {shap}
\bibfield{author}{\bibinfo{person}{Scott~M Lundberg} {and} \bibinfo{person}{Su-In Lee}.} \bibinfo{year}{2017}\natexlab{}.
\newblock \showarticletitle{A Unified Approach to Interpreting Model Predictions}.
\newblock In \bibinfo{booktitle}{\emph{Advances in Neural Information Processing Systems 30}}, \bibfield{editor}{\bibinfo{person}{I.~Guyon}, \bibinfo{person}{U.~V. Luxburg}, \bibinfo{person}{S.~Bengio}, \bibinfo{person}{H.~Wallach}, \bibinfo{person}{R.~Fergus}, \bibinfo{person}{S.~Vishwanathan}, {and} \bibinfo{person}{R.~Garnett}} (Eds.). \bibinfo{publisher}{Curran Associates, Inc.}, \bibinfo{pages}{4765--4774}.
\newblock
\urldef\tempurl%
\url{http://papers.nips.cc/paper/7062-a-unified-approach-to-interpreting-model-predictions.pdf}
\showURL{%
\tempurl}


\bibitem[MacGlashan et~al\mbox{.}(2017)]%
        {macglashan2017interactive}
\bibfield{author}{\bibinfo{person}{James MacGlashan}, \bibinfo{person}{Mark~K Ho}, \bibinfo{person}{Robert Loftin}, \bibinfo{person}{Bei Peng}, \bibinfo{person}{Guan Wang}, \bibinfo{person}{David~L Roberts}, \bibinfo{person}{Matthew~E Taylor}, {and} \bibinfo{person}{Michael~L Littman}.} \bibinfo{year}{2017}\natexlab{}.
\newblock \showarticletitle{Interactive learning from policy-dependent human feedback}. In \bibinfo{booktitle}{\emph{International conference on machine learning}}. PMLR, \bibinfo{pages}{2285--2294}.
\newblock


\bibitem[Manz({[n.\,d.]})]%
        {Manz_anywidget}
\bibfield{author}{\bibinfo{person}{Trevor Manz}.} \bibinfo{year}{[n.\,d.]}\natexlab{}.
\newblock \bibinfo{booktitle}{\emph{{anywidget}}}.
\newblock
\urldef\tempurl%
\url{https://github.com/manzt/anywidget}
\showURL{%
\tempurl}


\bibitem[Mark et~al\mbox{.}(2008)]%
        {mark2008cost}
\bibfield{author}{\bibinfo{person}{Gloria Mark}, \bibinfo{person}{Daniela Gudith}, {and} \bibinfo{person}{Ulrich Klocke}.} \bibinfo{year}{2008}\natexlab{}.
\newblock \showarticletitle{The cost of interrupted work: more speed and stress}. In \bibinfo{booktitle}{\emph{Proceedings of the SIGCHI conference on Human Factors in Computing Systems}}. \bibinfo{pages}{107--110}.
\newblock


\bibitem[McCurdy et~al\mbox{.}(2016)]%
        {mccurdy2016action}
\bibfield{author}{\bibinfo{person}{Nina McCurdy}, \bibinfo{person}{Jason Dykes}, {and} \bibinfo{person}{Miriah Meyer}.} \bibinfo{year}{2016}\natexlab{}.
\newblock \showarticletitle{Action design research and visualization design}. In \bibinfo{booktitle}{\emph{Proceedings of the Sixth Workshop on Beyond Time and Errors on Novel Evaluation Methods for Visualization}}. \bibinfo{pages}{10--18}.
\newblock


\bibitem[McInnes et~al\mbox{.}(2018)]%
        {mcinnes2018umapsoftware}
\bibfield{author}{\bibinfo{person}{Leland McInnes}, \bibinfo{person}{John Healy}, \bibinfo{person}{Nathaniel Saul}, {and} \bibinfo{person}{Lukas Grossberger}.} \bibinfo{year}{2018}\natexlab{}.
\newblock \showarticletitle{UMAP: Uniform Manifold Approximation and Projection}.
\newblock \bibinfo{journal}{\emph{The Journal of Open Source Software}} \bibinfo{volume}{3}, \bibinfo{number}{29} (\bibinfo{year}{2018}), \bibinfo{pages}{861}.
\newblock


\bibitem[McKenna et~al\mbox{.}(2014)]%
        {mckenna2014design}
\bibfield{author}{\bibinfo{person}{Sean McKenna}, \bibinfo{person}{Dominika Mazur}, \bibinfo{person}{James Agutter}, {and} \bibinfo{person}{Miriah Meyer}.} \bibinfo{year}{2014}\natexlab{}.
\newblock \showarticletitle{Design activity framework for visualization design}.
\newblock \bibinfo{journal}{\emph{IEEE Transactions on Visualization and Computer Graphics}} \bibinfo{volume}{20}, \bibinfo{number}{12} (\bibinfo{year}{2014}), \bibinfo{pages}{2191--2200}.
\newblock


\bibitem[Miller(2019)]%
        {miller2019explanation}
\bibfield{author}{\bibinfo{person}{Tim Miller}.} \bibinfo{year}{2019}\natexlab{}.
\newblock \showarticletitle{Explanation in artificial intelligence: Insights from the social sciences}.
\newblock \bibinfo{journal}{\emph{Artificial intelligence}}  \bibinfo{volume}{267} (\bibinfo{year}{2019}), \bibinfo{pages}{1--38}.
\newblock


\bibitem[Nori et~al\mbox{.}(2019)]%
        {nori2019interpretml}
\bibfield{author}{\bibinfo{person}{Harsha Nori}, \bibinfo{person}{Samuel Jenkins}, \bibinfo{person}{Paul Koch}, {and} \bibinfo{person}{Rich Caruana}.} \bibinfo{year}{2019}\natexlab{}.
\newblock \showarticletitle{InterpretML: A Unified Framework for Machine Learning Interpretability}.
\newblock \bibinfo{journal}{\emph{arXiv preprint arXiv:1909.09223}} (\bibinfo{year}{2019}).
\newblock


\bibitem[Paszke et~al\mbox{.}(2019)]%
        {PaszkePyTorchAnImperative2019}
\bibfield{author}{\bibinfo{person}{Adam Paszke}, \bibinfo{person}{Sam Gross}, \bibinfo{person}{Francisco Massa}, \bibinfo{person}{Adam Lerer}, \bibinfo{person}{James Bradbury}, \bibinfo{person}{Gregory Chanan}, \bibinfo{person}{Trevor Killeen}, \bibinfo{person}{Zeming Lin}, \bibinfo{person}{Natalia Gimelshein}, \bibinfo{person}{Luca Antiga}, \bibinfo{person}{Alban Desmaison}, \bibinfo{person}{Andreas Kopf}, \bibinfo{person}{Edward Yang}, \bibinfo{person}{Zachary DeVito}, \bibinfo{person}{Martin Raison}, \bibinfo{person}{Alykhan Tejani}, \bibinfo{person}{Sasank Chilamkurthy}, \bibinfo{person}{Benoit Steiner}, \bibinfo{person}{Lu Fang}, \bibinfo{person}{Junjie Bai}, {and} \bibinfo{person}{Soumith Chintala}.} \bibinfo{year}{2019}\natexlab{}.
\newblock \showarticletitle{{PyTorch: An Imperative Style, High-Performance Deep Learning Library}}. In \bibinfo{booktitle}{\emph{Advances in Neural Information Processing Systems 32}}, \bibfield{editor}{\bibinfo{person}{H.~Wallach}, \bibinfo{person}{H.~Larochelle}, \bibinfo{person}{A.~Beygelzimer}, \bibinfo{person}{F.~d'Alché Buc}, \bibinfo{person}{E.~Fox}, {and} \bibinfo{person}{R.~Garnett}} (Eds.). \bibinfo{publisher}{Curran Associates, Inc.}, \bibinfo{pages}{8024--8035}.
\newblock
\urldef\tempurl%
\url{http://papers.neurips.cc/paper/9015-pytorch-an-imperative-style-high-performance-deep-learning-library.pdf}
\showURL{%
\tempurl}


\bibitem[Pedregosa et~al\mbox{.}(2011)]%
        {scikitlearn}
\bibfield{author}{\bibinfo{person}{F. Pedregosa}, \bibinfo{person}{G. Varoquaux}, \bibinfo{person}{A. Gramfort}, \bibinfo{person}{V. Michel}, \bibinfo{person}{B. Thirion}, \bibinfo{person}{O. Grisel}, \bibinfo{person}{M. Blondel}, \bibinfo{person}{P. Prettenhofer}, \bibinfo{person}{R. Weiss}, \bibinfo{person}{V. Dubourg}, \bibinfo{person}{J. Vanderplas}, \bibinfo{person}{A. Passos}, \bibinfo{person}{D. Cournapeau}, \bibinfo{person}{M. Brucher}, \bibinfo{person}{M. Perrot}, {and} \bibinfo{person}{E. Duchesnay}.} \bibinfo{year}{2011}\natexlab{}.
\newblock \showarticletitle{Scikit-learn: Machine Learning in {P}ython}.
\newblock \bibinfo{journal}{\emph{Journal of Machine Learning Research}}  \bibinfo{volume}{12} (\bibinfo{year}{2011}), \bibinfo{pages}{2825--2830}.
\newblock


\bibitem[Pike et~al\mbox{.}(2009)]%
        {pike2009science}
\bibfield{author}{\bibinfo{person}{William~A Pike}, \bibinfo{person}{John Stasko}, \bibinfo{person}{Remco Chang}, {and} \bibinfo{person}{Theresa~A O'connell}.} \bibinfo{year}{2009}\natexlab{}.
\newblock \showarticletitle{The science of interaction}.
\newblock \bibinfo{journal}{\emph{Information visualization}} \bibinfo{volume}{8}, \bibinfo{number}{4} (\bibinfo{year}{2009}), \bibinfo{pages}{263--274}.
\newblock


\bibitem[Raghunandan et~al\mbox{.}(2024)]%
        {raghunandan2024lodestar}
\bibfield{author}{\bibinfo{person}{Deepthi Raghunandan}, \bibinfo{person}{Zhe Cui}, \bibinfo{person}{Kartik Krishnan}, \bibinfo{person}{Segen Tirfe}, \bibinfo{person}{Shenzhi Shi}, \bibinfo{person}{Tejaswi~Darshan Shrestha}, \bibinfo{person}{Leilani Battle}, {and} \bibinfo{person}{Niklas Elmqvist}.} \bibinfo{year}{2024}\natexlab{}.
\newblock \showarticletitle{Lodestar: Supporting rapid prototyping of data science workflows through data-driven analysis recommendations}.
\newblock \bibinfo{journal}{\emph{Information Visualization}} \bibinfo{volume}{23}, \bibinfo{number}{1} (\bibinfo{year}{2024}), \bibinfo{pages}{21--39}.
\newblock


\bibitem[Ribeiro et~al\mbox{.}(2016)]%
        {lime}
\bibfield{author}{\bibinfo{person}{Marco~Tulio Ribeiro}, \bibinfo{person}{Sameer Singh}, {and} \bibinfo{person}{Carlos Guestrin}.} \bibinfo{year}{2016}\natexlab{}.
\newblock \showarticletitle{"Why Should {I} Trust You?": Explaining the Predictions of Any Classifier}. In \bibinfo{booktitle}{\emph{Proceedings of the 22nd {ACM} {SIGKDD} International Conference on Knowledge Discovery and Data Mining, San Francisco, CA, USA, August 13-17, 2016}}. \bibinfo{pages}{1135--1144}.
\newblock


\bibitem[Roh et~al\mbox{.}(2019)]%
        {roh2019survey}
\bibfield{author}{\bibinfo{person}{Yuji Roh}, \bibinfo{person}{Geon Heo}, {and} \bibinfo{person}{Steven~Euijong Whang}.} \bibinfo{year}{2019}\natexlab{}.
\newblock \showarticletitle{A survey on data collection for machine learning: a big data-ai integration perspective}.
\newblock \bibinfo{journal}{\emph{IEEE Transactions on Knowledge and Data Engineering}} \bibinfo{volume}{33}, \bibinfo{number}{4} (\bibinfo{year}{2019}), \bibinfo{pages}{1328--1347}.
\newblock


\bibitem[Ross et~al\mbox{.}(2021)]%
        {ross2021evaluating}
\bibfield{author}{\bibinfo{person}{Andrew Ross}, \bibinfo{person}{Nina Chen}, \bibinfo{person}{Elisa~Zhao Hang}, \bibinfo{person}{Elena~L Glassman}, {and} \bibinfo{person}{Finale Doshi-Velez}.} \bibinfo{year}{2021}\natexlab{}.
\newblock \showarticletitle{Evaluating the interpretability of generative models by interactive reconstruction}. In \bibinfo{booktitle}{\emph{Proceedings of the 2021 CHI Conference on Human Factors in Computing Systems}}. \bibinfo{pages}{1--15}.
\newblock


\bibitem[Rouzrokh et~al\mbox{.}(2023)]%
        {rouzrokh_machine_2023}
\bibfield{author}{\bibinfo{person}{Pouria Rouzrokh}, \bibinfo{person}{Bardia Khosravi}, \bibinfo{person}{Sanaz Vahdati}, \bibinfo{person}{Mana Moassefi}, \bibinfo{person}{Shahriar Faghani}, \bibinfo{person}{Elham Mahmoudi}, \bibinfo{person}{Hamid Chalian}, {and} \bibinfo{person}{Bradley~J. Erickson}.} \bibinfo{year}{2023}\natexlab{}.
\newblock \showarticletitle{Machine {Learning} in {Cardiovascular} {Imaging}: {A} {Scoping} {Review} of {Published} {Literature}}.
\newblock \bibinfo{journal}{\emph{Current Radiology Reports}} \bibinfo{volume}{11}, \bibinfo{number}{2} (\bibinfo{date}{Feb.} \bibinfo{year}{2023}), \bibinfo{pages}{34--45}.
\newblock
\showISSN{2167-4825}
\urldef\tempurl%
\url{https://doi.org/10.1007/s40134-022-00407-8}
\showDOI{\tempurl}


\bibitem[Schwab et~al\mbox{.}(2000)]%
        {schwab2000making}
\bibfield{author}{\bibinfo{person}{Matthias Schwab}, \bibinfo{person}{N Karrenbach}, {and} \bibinfo{person}{Jon Claerbout}.} \bibinfo{year}{2000}\natexlab{}.
\newblock \showarticletitle{Making scientific computations reproducible}.
\newblock \bibinfo{journal}{\emph{Computing in Science \& Engineering}} \bibinfo{volume}{2}, \bibinfo{number}{6} (\bibinfo{year}{2000}), \bibinfo{pages}{61--67}.
\newblock


\bibitem[Sedlmair et~al\mbox{.}(2012)]%
        {sedlmair2012design}
\bibfield{author}{\bibinfo{person}{Michael Sedlmair}, \bibinfo{person}{Miriah Meyer}, {and} \bibinfo{person}{Tamara Munzner}.} \bibinfo{year}{2012}\natexlab{}.
\newblock \showarticletitle{Design study methodology: Reflections from the trenches and the stacks}.
\newblock \bibinfo{journal}{\emph{IEEE transactions on visualization and computer graphics}} \bibinfo{volume}{18}, \bibinfo{number}{12} (\bibinfo{year}{2012}), \bibinfo{pages}{2431--2440}.
\newblock


\bibitem[Sermesant et~al\mbox{.}(2021)]%
        {sermesant_applications_2021}
\bibfield{author}{\bibinfo{person}{Maxime Sermesant}, \bibinfo{person}{Hervé Delingette}, \bibinfo{person}{Hubert Cochet}, \bibinfo{person}{Pierre Jaïs}, {and} \bibinfo{person}{Nicholas Ayache}.} \bibinfo{year}{2021}\natexlab{}.
\newblock \showarticletitle{Applications of artificial intelligence in cardiovascular imaging}.
\newblock \bibinfo{journal}{\emph{Nature Reviews Cardiology}} \bibinfo{volume}{18}, \bibinfo{number}{8} (\bibinfo{date}{Aug.} \bibinfo{year}{2021}), \bibinfo{pages}{600--609}.
\newblock
\showISSN{1759-5010}
\urldef\tempurl%
\url{https://doi.org/10.1038/s41569-021-00527-2}
\showDOI{\tempurl}


\bibitem[Shen(2014)]%
        {shen2014interactive}
\bibfield{author}{\bibinfo{person}{Helen Shen}.} \bibinfo{year}{2014}\natexlab{}.
\newblock \showarticletitle{Interactive notebooks: Sharing the code}.
\newblock \bibinfo{journal}{\emph{Nature}} \bibinfo{volume}{515}, \bibinfo{number}{7525} (\bibinfo{year}{2014}), \bibinfo{pages}{152--152}.
\newblock


\bibitem[Sovrano and Vitali(2021)]%
        {sovrano2021philosophy}
\bibfield{author}{\bibinfo{person}{Francesco Sovrano} {and} \bibinfo{person}{Fabio Vitali}.} \bibinfo{year}{2021}\natexlab{}.
\newblock \showarticletitle{From Philosophy to Interfaces: An Explanatory Method and a Tool Inspired by Achinstein’s Theory of Explanation}. In \bibinfo{booktitle}{\emph{26th International Conference on Intelligent User Interfaces}}. \bibinfo{pages}{81--91}.
\newblock


\bibitem[Sperrle et~al\mbox{.}(2021)]%
        {sperrle2021survey}
\bibfield{author}{\bibinfo{person}{Fabian Sperrle}, \bibinfo{person}{Mennatallah El-Assady}, \bibinfo{person}{Grace Guo}, \bibinfo{person}{Rita Borgo}, \bibinfo{person}{D~Horng Chau}, \bibinfo{person}{Alex Endert}, {and} \bibinfo{person}{Daniel Keim}.} \bibinfo{year}{2021}\natexlab{}.
\newblock \showarticletitle{A Survey of Human-Centered Evaluations in Human-Centered Machine Learning}. In \bibinfo{booktitle}{\emph{Computer Graphics Forum}}, Vol.~\bibinfo{volume}{40}. Wiley Online Library, \bibinfo{pages}{543--568}.
\newblock


\bibitem[Spinner et~al\mbox{.}(2019)]%
        {spinner2019explainer}
\bibfield{author}{\bibinfo{person}{Thilo Spinner}, \bibinfo{person}{Udo Schlegel}, \bibinfo{person}{Hanna Sch{\"a}fer}, {and} \bibinfo{person}{Mennatallah El-Assady}.} \bibinfo{year}{2019}\natexlab{}.
\newblock \showarticletitle{explAIner: A visual analytics framework for interactive and explainable machine learning}.
\newblock \bibinfo{journal}{\emph{IEEE transactions on visualization and computer graphics}} \bibinfo{volume}{26}, \bibinfo{number}{1} (\bibinfo{year}{2019}), \bibinfo{pages}{1064--1074}.
\newblock


\bibitem[Syeda et~al\mbox{.}(2020)]%
        {syeda2020design}
\bibfield{author}{\bibinfo{person}{Uzma~Haque Syeda}, \bibinfo{person}{Prasanth Murali}, \bibinfo{person}{Lisa Roe}, \bibinfo{person}{Becca Berkey}, {and} \bibinfo{person}{Michelle~A Borkin}.} \bibinfo{year}{2020}\natexlab{}.
\newblock \showarticletitle{Design Study" Lite" Methodology: Expediting Design Studies and Enabling the Synergy of Visualization Pedagogy and Social Good}. In \bibinfo{booktitle}{\emph{Proceedings of the 2020 CHI Conference on Human Factors in Computing Systems}}. \bibinfo{pages}{1--13}.
\newblock


\bibitem[van Assen et~al\mbox{.}(2023)]%
        {van_assen_artificial_2023}
\bibfield{author}{\bibinfo{person}{Marly van Assen}, \bibinfo{person}{Alexander~C Razavi}, \bibinfo{person}{Seamus~P Whelton}, {and} \bibinfo{person}{Carlo~N De~Cecco}.} \bibinfo{year}{2023}\natexlab{}.
\newblock \showarticletitle{Artificial intelligence in cardiac imaging: where we are and what we want}.
\newblock \bibinfo{journal}{\emph{European Heart Journal}} \bibinfo{volume}{44}, \bibinfo{number}{7} (\bibinfo{date}{Feb.} \bibinfo{year}{2023}), \bibinfo{pages}{541--543}.
\newblock
\showISSN{0195-668X}
\urldef\tempurl%
\url{https://doi.org/10.1093/eurheartj/ehac700}
\showDOI{\tempurl}


\bibitem[VanderPlas et~al\mbox{.}(2018)]%
        {VanderPlas2018}
\bibfield{author}{\bibinfo{person}{Jacob VanderPlas}, \bibinfo{person}{Brian Granger}, \bibinfo{person}{Jeffrey Heer}, \bibinfo{person}{Dominik Moritz}, \bibinfo{person}{Kanit Wongsuphasawat}, \bibinfo{person}{Arvind Satyanarayan}, \bibinfo{person}{Eitan Lees}, \bibinfo{person}{Ilia Timofeev}, \bibinfo{person}{Ben Welsh}, {and} \bibinfo{person}{Scott Sievert}.} \bibinfo{year}{2018}\natexlab{}.
\newblock \showarticletitle{Altair: Interactive Statistical Visualizations for Python}.
\newblock \bibinfo{journal}{\emph{Journal of Open Source Software}} \bibinfo{volume}{3}, \bibinfo{number}{32} (\bibinfo{year}{2018}), \bibinfo{pages}{1057}.
\newblock
\urldef\tempurl%
\url{https://doi.org/10.21105/joss.01057}
\showDOI{\tempurl}


\bibitem[Wadsworth(1993)]%
        {wadsworth1993participatory}
\bibfield{author}{\bibinfo{person}{Yoland Wadsworth}.} \bibinfo{year}{1993}\natexlab{}.
\newblock \bibinfo{booktitle}{\emph{What is participatory action research?}}
\newblock \bibinfo{publisher}{Action Research Issues Association}.
\newblock


\bibitem[Wang et~al\mbox{.}(2023)]%
        {wang2023slide4n}
\bibfield{author}{\bibinfo{person}{Fengjie Wang}, \bibinfo{person}{Xuye Liu}, \bibinfo{person}{Oujing Liu}, \bibinfo{person}{Ali Neshati}, \bibinfo{person}{Tengfei Ma}, \bibinfo{person}{Min Zhu}, {and} \bibinfo{person}{Jian Zhao}.} \bibinfo{year}{2023}\natexlab{}.
\newblock \showarticletitle{Slide4N: Creating Presentation Slides from Computational Notebooks with Human-AI Collaboration}. In \bibinfo{booktitle}{\emph{Proceedings of the 2023 CHI Conference on Human Factors in Computing Systems}}. \bibinfo{pages}{1--18}.
\newblock


\bibitem[Wang et~al\mbox{.}(2022a)]%
        {wang2022stickyland}
\bibfield{author}{\bibinfo{person}{Zijie~J Wang}, \bibinfo{person}{Katie Dai}, {and} \bibinfo{person}{W~Keith Edwards}.} \bibinfo{year}{2022}\natexlab{a}.
\newblock \showarticletitle{Stickyland: Breaking the linear presentation of computational notebooks}. In \bibinfo{booktitle}{\emph{CHI Conference on Human Factors in Computing Systems Extended Abstracts}}. \bibinfo{pages}{1--7}.
\newblock


\bibitem[Wang et~al\mbox{.}(2022b)]%
        {wang2022nova}
\bibfield{author}{\bibinfo{person}{Zijie~J Wang}, \bibinfo{person}{David Munechika}, \bibinfo{person}{Seongmin Lee}, {and} \bibinfo{person}{Duen~Horng Chau}.} \bibinfo{year}{2022}\natexlab{b}.
\newblock \showarticletitle{Nova: A practical method for creating notebook-ready visual analytics}.
\newblock \bibinfo{journal}{\emph{arXiv preprint arXiv:2205.03963}} (\bibinfo{year}{2022}).
\newblock


\bibitem[Warnell et~al\mbox{.}(2018)]%
        {warnell2018deep}
\bibfield{author}{\bibinfo{person}{Garrett Warnell}, \bibinfo{person}{Nicholas Waytowich}, \bibinfo{person}{Vernon Lawhern}, {and} \bibinfo{person}{Peter Stone}.} \bibinfo{year}{2018}\natexlab{}.
\newblock \showarticletitle{Deep tamer: Interactive agent shaping in high-dimensional state spaces}. In \bibinfo{booktitle}{\emph{Proceedings of the AAAI conference on artificial intelligence}}, Vol.~\bibinfo{volume}{32}.
\newblock


\bibitem[Waskom(2021)]%
        {Waskom2021}
\bibfield{author}{\bibinfo{person}{Michael~L. Waskom}.} \bibinfo{year}{2021}\natexlab{}.
\newblock \showarticletitle{seaborn: statistical data visualization}.
\newblock \bibinfo{journal}{\emph{Journal of Open Source Software}} \bibinfo{volume}{6}, \bibinfo{number}{60} (\bibinfo{year}{2021}), \bibinfo{pages}{3021}.
\newblock
\urldef\tempurl%
\url{https://doi.org/10.21105/joss.03021}
\showDOI{\tempurl}


\bibitem[Weidinger et~al\mbox{.}(2021)]%
        {weidinger2021ethical}
\bibfield{author}{\bibinfo{person}{Laura Weidinger}, \bibinfo{person}{John Mellor}, \bibinfo{person}{Maribeth Rauh}, \bibinfo{person}{Conor Griffin}, \bibinfo{person}{Jonathan Uesato}, \bibinfo{person}{Po-Sen Huang}, \bibinfo{person}{Myra Cheng}, \bibinfo{person}{Mia Glaese}, \bibinfo{person}{Borja Balle}, \bibinfo{person}{Atoosa Kasirzadeh}, {et~al\mbox{.}}} \bibinfo{year}{2021}\natexlab{}.
\newblock \showarticletitle{Ethical and social risks of harm from language models}.
\newblock \bibinfo{journal}{\emph{arXiv preprint arXiv:2112.04359}} (\bibinfo{year}{2021}).
\newblock


\bibitem[Weinman et~al\mbox{.}(2021)]%
        {weinman2021fork}
\bibfield{author}{\bibinfo{person}{Nathaniel Weinman}, \bibinfo{person}{Steven~M Drucker}, \bibinfo{person}{Titus Barik}, {and} \bibinfo{person}{Robert DeLine}.} \bibinfo{year}{2021}\natexlab{}.
\newblock \showarticletitle{Fork it: Supporting stateful alternatives in computational notebooks}. In \bibinfo{booktitle}{\emph{Proceedings of the 2021 CHI Conference on Human Factors in Computing Systems}}. \bibinfo{pages}{1--12}.
\newblock


\bibitem[Whang and Lee(2020)]%
        {whang2020data}
\bibfield{author}{\bibinfo{person}{Steven~Euijong Whang} {and} \bibinfo{person}{Jae-Gil Lee}.} \bibinfo{year}{2020}\natexlab{}.
\newblock \showarticletitle{Data collection and quality challenges for deep learning}.
\newblock \bibinfo{journal}{\emph{Proceedings of the VLDB Endowment}} \bibinfo{volume}{13}, \bibinfo{number}{12} (\bibinfo{year}{2020}), \bibinfo{pages}{3429--3432}.
\newblock


\bibitem[Whang et~al\mbox{.}(2023)]%
        {whang2023data}
\bibfield{author}{\bibinfo{person}{Steven~Euijong Whang}, \bibinfo{person}{Yuji Roh}, \bibinfo{person}{Hwanjun Song}, {and} \bibinfo{person}{Jae-Gil Lee}.} \bibinfo{year}{2023}\natexlab{}.
\newblock \showarticletitle{Data collection and quality challenges in deep learning: A data-centric ai perspective}.
\newblock \bibinfo{journal}{\emph{The VLDB Journal}} \bibinfo{volume}{32}, \bibinfo{number}{4} (\bibinfo{year}{2023}), \bibinfo{pages}{791--813}.
\newblock


\bibitem[widgets community(2023)]%
        {JupyterWidgets}
\bibfield{author}{\bibinfo{person}{Jupyter widgets community}.} \bibinfo{year}{2023}\natexlab{}.
\newblock \bibinfo{title}{Jupyter Widgets}.
\newblock
\newblock
\urldef\tempurl%
\url{https://ipywidgets.readthedocs.io/en/stable/}
\showURL{%
\tempurl}
\newblock
\shownote{Accessed Feb 20, 2024}.


\bibitem[Wirth et~al\mbox{.}(2017)]%
        {wirth2017survey}
\bibfield{author}{\bibinfo{person}{Christian Wirth}, \bibinfo{person}{Riad Akrour}, \bibinfo{person}{Gerhard Neumann}, \bibinfo{person}{Johannes F{\"u}rnkranz}, {et~al\mbox{.}}} \bibinfo{year}{2017}\natexlab{}.
\newblock \showarticletitle{A survey of preference-based reinforcement learning methods}.
\newblock \bibinfo{journal}{\emph{Journal of Machine Learning Research}} \bibinfo{volume}{18}, \bibinfo{number}{136} (\bibinfo{year}{2017}), \bibinfo{pages}{1--46}.
\newblock


\bibitem[Wu et~al\mbox{.}(2020)]%
        {wu2020b2}
\bibfield{author}{\bibinfo{person}{Yifan Wu}, \bibinfo{person}{Joseph~M Hellerstein}, {and} \bibinfo{person}{Arvind Satyanarayan}.} \bibinfo{year}{2020}\natexlab{}.
\newblock \showarticletitle{B2: Bridging code and interactive visualization in computational notebooks}. In \bibinfo{booktitle}{\emph{Proceedings of the 33rd Annual ACM Symposium on User Interface Software and Technology}}. \bibinfo{pages}{152--165}.
\newblock


\bibitem[Yang et~al\mbox{.}(2022)]%
        {wenzhuo2022omnixai}
\bibfield{author}{\bibinfo{person}{Wenzhuo Yang}, \bibinfo{person}{Hung Le}, \bibinfo{person}{Silvio Savarese}, {and} \bibinfo{person}{Steven Hoi}.} \bibinfo{year}{2022}\natexlab{}.
\newblock \showarticletitle{OmniXAI: A Library for Explainable AI}.
\newblock  (\bibinfo{year}{2022}).
\newblock
\urldef\tempurl%
\url{https://doi.org/10.48550/ARXIV.2206.01612}
\showDOI{\tempurl}
\showeprint[arxiv]{206.01612}


\bibitem[Zheng et~al\mbox{.}(2022)]%
        {zheng2022telling}
\bibfield{author}{\bibinfo{person}{Chengbo Zheng}, \bibinfo{person}{Dakuo Wang}, \bibinfo{person}{April~Yi Wang}, {and} \bibinfo{person}{Xiaojuan Ma}.} \bibinfo{year}{2022}\natexlab{}.
\newblock \showarticletitle{Telling stories from computational notebooks: Ai-assisted presentation slides creation for presenting data science work}. In \bibinfo{booktitle}{\emph{Proceedings of the 2022 CHI Conference on Human Factors in Computing Systems}}. \bibinfo{pages}{1--20}.
\newblock


\end{thebibliography}

\end{document}